\documentclass[a4paper]{IEEEtran}

\usepackage[latin1]{inputenc}
\usepackage{amsmath}
\usepackage{amssymb}
\usepackage{epsfig}
\usepackage{xspace}
\usepackage[bf,footnotesize]{caption2}
\usepackage{pstricks}
\usepackage{psfrag}
\usepackage{color}
\usepackage{rotating}

\usepackage{citesort}

\usepackage{mathptmx}

\newcommand{\eq}[1]      {(\ref{#1})}

\newcommand{\Fig}[1]     {Fig.~\ref{#1}}

\newcommand{\Sec}[1]     {Section~\ref{#1}}

\newcommand{\PD}[2]     {\frac{\partial #1}{\partial #2}}

\newcommand{\Vth}       {V_\mathrm{th}}

\newcommand{\dVth}      {\Delta \Vth}

\newcommand{\VDD}       {V_\mathrm{DD}}

\newcommand{\statav}   [1]{\ensuremath{\langle #1 \rangle}}

\newcommand{\taue}      {\tau_\nrg}

\newcommand{\q}         {{\mathrm{q}}}

\newcommand{\Goes}      {Goes}

\newcommand{\degC}        {\ensuremath{\,^\circ\text{C}}}
\newcommand{\U}[2]        {\ensuremath{#1\,\mathrm{#2}}}

\newcounter{myitem}
\newcommand{\mycnt}  {{\it (\roman{myitem})}\addtocounter{myitem}{1}}

\newcommand{\resetmycnt}{\setcounter{myitem}{1}}

\resetmycnt

\newcommand{\LFig}  {{\bf Left}}

\newcommand{\RFig}  {{\bf Right}}
\newcommand{\TFig}  {{\bf Top}}
\newcommand{\BFig}  {{\bf Bottom}}

\newcommand{\Hyd}         {\ensuremath{\mathrm{H}}}

\newcommand{\Hydt}        {\ensuremath{\mathrm{H_2}}}

\newcommand{\Si}          {\ensuremath{\mathrm{Si}}}

\newcommand{\SiOt}        {\ensuremath{\mathrm{SiO_2}}}

\newcommand{\HfOt}        {\ensuremath{\mathrm{HfO_2}}}

\newcommand{\tox}         {t_\mathrm{ox}}

\newcommand{\tstress}     {t_\mathrm{s}}
\newcommand{\trelax}      {t_\mathrm{r}}

\newcommand{\Nit}         {N_\mathrm{it}}

\newcommand{\Not}         {N_\mathrm{ot}}

\newcommand{\Pb}          {\ensuremath{P_\mathrm{b}}}

\newcommand{\Pbz}         {\ensuremath{P_\mathrm{b0}}}
\newcommand{\Pbo}         {\ensuremath{P_\mathrm{b1}}}

\renewcommand{\taue}      {\tau_\mathrm{e}}
\newcommand{\atauc}       {\bar{\tau}_\mathrm{c}}
\newcommand{\ataue}       {\bar{\tau}_\mathrm{e}}

\newcommand{\bondone}     {\text{--}}

\newcommand{\Vstress} {V_\mathrm{stress}}

\newcommand{\pdf}  {p.d.f.}

\newcommand{\epsz} {\epsilon_0}
\newcommand{\epsr} {\epsilon_\mathrm{r}}

\newcommand{\fRD}   {f_\mathrm{RD}}

\newcommand{\atauci}  {\bar{\tau}_{\mathrm{c},i}}
\newcommand{\atauei}  {\bar{\tau}_{\mathrm{e},i}}
\newcommand{\feta}  {f_\eta}
\newcommand{\etai}  {\eta_i}
\newcommand{\setai}  {\sigma_{\eta,i}}
\newcommand{\aeta}  {\bar{\eta}}
\newcommand{\aetar}  {\bar{\eta_\mathrm{r}}}
\newcommand{\aetai}  {\bar{\eta}_i}

\renewcommand{\taue}  {\tau_\mathrm{e}}

\newcommand{\etar}    {\eta_\mathrm{r}}
\newcommand{\etaz}    {\eta_\mathrm{0}}

\newcommand{\figwidth}{5.2cm}
\newcommand{\sfigwidth}{4.cm}
\newcommand{\mapwidth}{4.3cm}

\newcommand{\capspace}{\vspace*{0cm}}
\newcommand{\figspace}{\vspace*{-0.5cm}}

\newcommand{\pRD}{poly \Hyd/\Hydt{}}

\renewcommand{\Vstress} {V_\mathrm{s}}

\begin{document}
\title{
{ NBTI in Nanoscale MOSFETs --\\ The Ultimate Modeling Benchmark}
}
\author{%
T. Grasser$^*$, 
K. Rott$^\bullet$, 
H. Reisinger$^\bullet$, 
M. Waltl$^*$, 
F. Schanovsky$^*$,
and
B. Kaczer$^\circ$ \\
\small %
   $^*$Institute for Microelectronics,
   TU Wien, Austria\hspace*{4mm}
   $^\bullet$ Infineon, Munich, Germany\hspace*{4mm}
   $^\circ$ IMEC, Leuven, Belgium\hspace*{4mm}
\\
\vspace*{-4mm}
}

\maketitle

\vspace*{-5mm}

\begin{abstract} 

After nearly half a century of research into the bias temperature
instability (BTI), two classes of models have emerged as the strongest
contenders: one class of models, the reaction-diffusion models, is
built around the idea that hydrogen is released from the interface and
that it is the \emph{diffusion} of some form of hydrogen that controls both
degradation and recovery. While many different variants of the reaction-diffusion idea have been published
over the years, the most commonly used recent models are based on non-dispersive
reaction rates and non-dispersive diffusion. The other class of models is based on the idea that
degradation is controlled by first-order \emph{reactions} with
widely distributed (dispersive) reaction rates.  
We demonstrate that 
\emph{these two classes give fundamentally different
predictions for the stochastic degradation and recovery of nanoscale devices, therefore providing the ultimate
modeling benchmark.}  Using detailed experimental time-dependent defect
spectroscopy (TDDS) data obtained on such nanoscale devices, we
investigate the compatibility of these models with experiment.
Our results show that the \emph{diffusion} of hydrogen (or any other species) is unlikely to be
the limiting aspect that determines degradation.  On the other hand,
the data are fully consistent with \emph{reaction}-limited
models.  We finally argue that only the correct understanding of the physical mechanisms
leading to the significant device-to-device variation observed in the degradation in nanoscale devices
will enable accurate reliability projections and device optimization.
\end{abstract}

\section{Introduction}

Research into the bias temperature instability (BTI) has revealed a
plethora of puzzling issues which have proven a formidable obstacle to
the understanding of the phenomenon
\cite{ALAM03,KACZER05,WANG06,REISINGER06,HUARD06,HOUSSA07,GRASSER07,ZHANG07B,HUARD10,REISINGER10,ANG11,GRASSER11B,ZOU12B,MAHAPATRA13}.
In particular, numerous modeling ideas have been put forward and
refined at various levels.  Most of these models have in common that
the overall degradation is assumed to be due to two components: one
component ($\Nit$) is related to the release of hydrogen from
passivated silicon dangling bonds at the interface, thereby forming
electrically active \Pb{} centers \cite{CAMPBELL07B}, while the other
($\Not$) is due to the trapping of holes in the oxide
\cite{ZHANG04,SHEN04,HUARD06,ZHANG07B,ANG08B,VEKSLER14}.  However, these models can differ significantly
in the details of the physical mechanisms invoked to explain the
degradation.

At present, from all these modeling attempts two classes have emerged
that appear to be able to explain a wide range of experimental
observations: the first class is built around the concept of the
reaction-diffusion (RD) model \cite{ALAM03,MAHAPATRA13}, where it is
assumed that it is the \emph{diffusion} of the released hydrogen that
dominates the dynamics.  The other class is based on the notion that
it is the \emph{reactions} which essentially limit the dynamics, and
that the reaction rates are distributed over a wide range
\cite{STESMANS00,HAGGAG01,HOUSSA05,HUARD06,GRASSER11F}.  In other words, in this \emph{reaction}-limited class of models, both
interface states ($\Nit$) and oxide charges ($\Not$) are assumed to be
(in the simplest case) created and annealed by first-order reactions.  In contrast, in
the \emph{diffusion}-limited class (RD models), the dynamics of $\Nit$ creation and annealing are assumed to be dominated
by a \emph{diffusion}-limited process, which controlles both long term degradation
and recovery.

Many of these models have been developed to such a high degree that
they appear to be able to predict a wide range of experimental
observations \cite{HUARD10,GRASSER10,GRASSER11B,MAHAPATRA13,GOEL14}.
Typically, however, experimental data are obtained on large-area
(macroscopic) devices where the microscopic physics are washed out by averaging.  In
nanoscale devices, on the other hand, it has been shown that the
creation and annihilation of individual defects can be observed at the
statistical level \cite{WANG06,HUARD06,REISINGER10,GRASSER10,ZOU12B}.
We will demonstrate in the following that \emph{this statistical information
provides the ultimate benchmark for any BTI model, as it reveals the
underlying microscopic physics to an unprecedented degree}.  This
allows for an evaluation of the foundations of the
two model classes, as it clearly answers the fundamental question:
\emph{is BTI reaction- or diffusion-limited}?   As such, the
benchmark provided here is simple and not clouded by the complexities of the
individual models.

\begin{figure}[!t]
\begin{center}
  \hbox{
    \includegraphics[width=\sfigwidth,angle=-90]{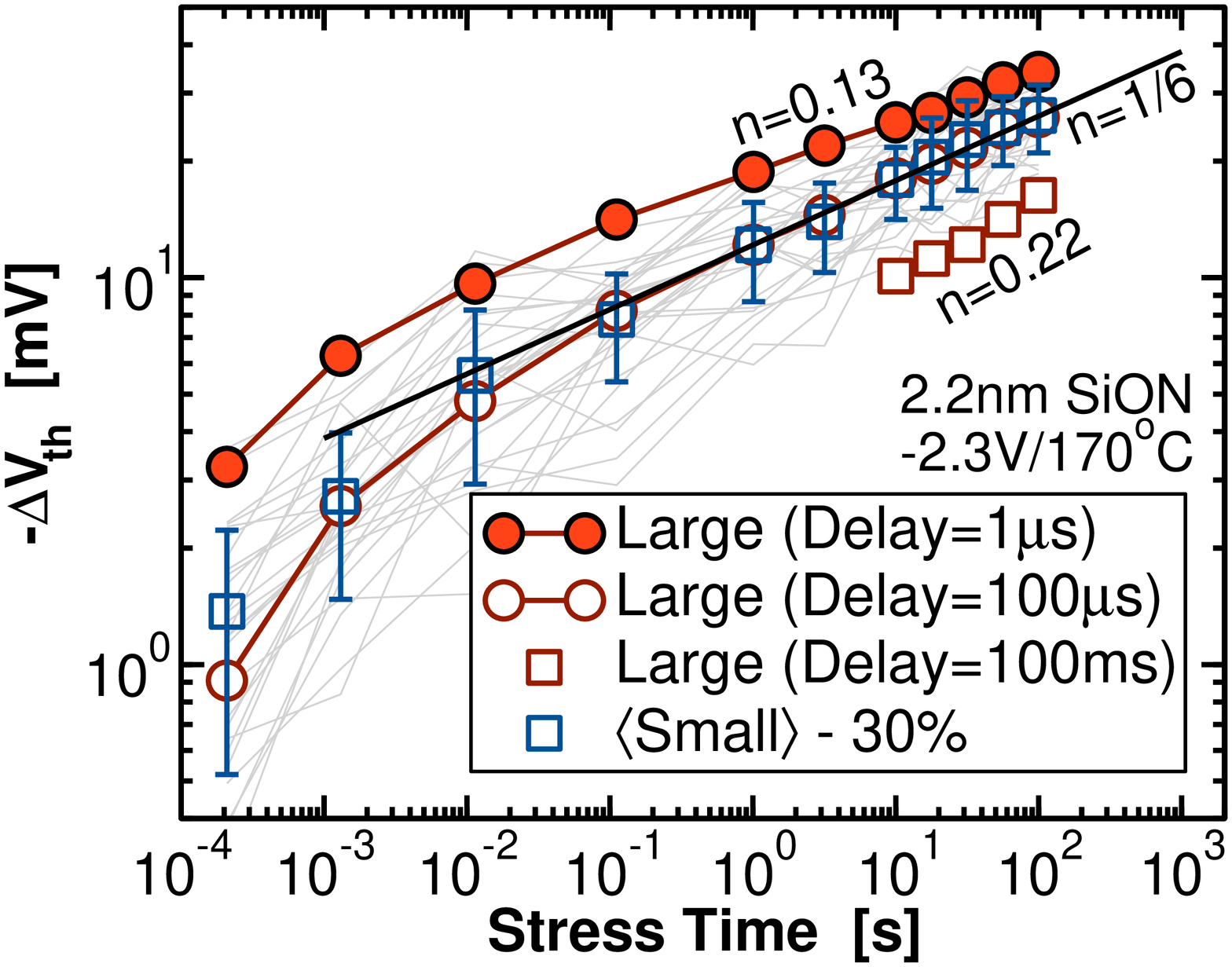}
    \hspace*{-4mm}
    \includegraphics[width=\sfigwidth,angle=-90]{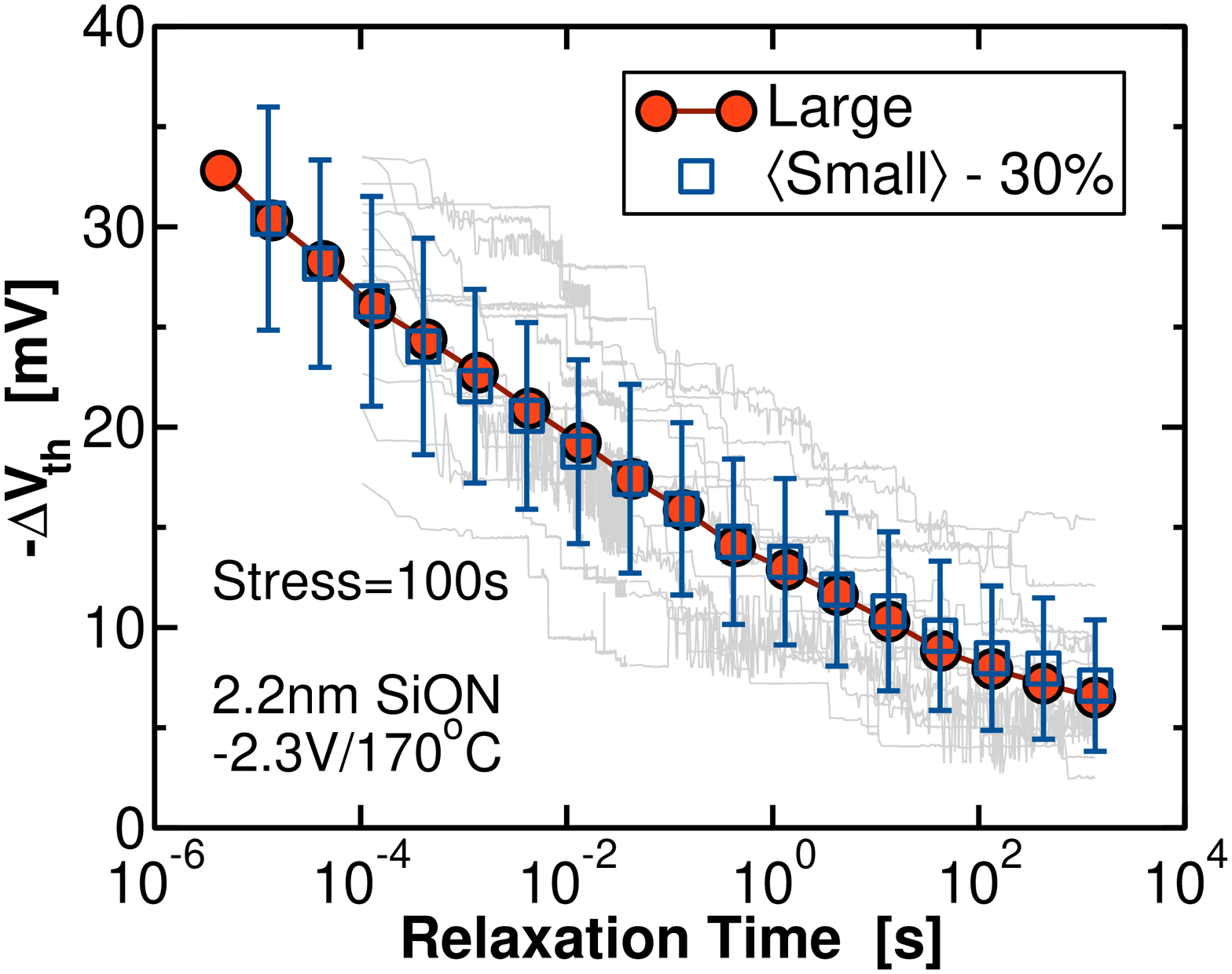}
  }
    \vspace{-6mm}

    \capspace\caption{\label{f:SmallLarge} Degradation (left) and recovery (right) of 27 small-area devices (light gray lines)
      ($\U{120}{nm} \times \U{280}{nm}$) compared to a large-area device (red symbols) with area $\U{120}{nm} \times \U{10}{\mu m}$.
      While the average degradation of the small-area devices is larger by 30\% (open symbols, error bars are $\pm \sigma$), the kinetics during both stress
      and recovery are otherwise identical.    
In particular, during
stress a power-law slope of $1/6$ is observed in both large and small-area
devices {\bf if} the measurement delay is chosen \U{100}{\mu s}.
      \vspace*{0\baselineskip}
    }
\end{center}
\figspace
\end{figure}

\section{Equivalence of Large and Small Devices}

Since the stochastic response of nanoscale devices to bias-temperature
stress lies at the heart of our arguments, we begin by experimentally
demonstrating the equivalence of large- and small-area devices.  For
this, we compare the degradation of a large-area device to the average
degradation observed in 27 small-area devices when subjected to
negative BTI (NBTI).  All measurements in the present study rely on the
  ultra-fast $\dVth$ technique published previously
  \cite{REISINGER06}, which has a delay of \U{1}{\mu s} on large devices.  Due to the lower
current levels, the delay increases to \U{100}{\mu s} in small-area devices.  As can be seen in \Fig{f:SmallLarge}, although
the degradation in small-area devices shows larger signs of
variability, discrete steps during recovery, and is about 30\% larger
than in this particular large-area device, the average dynamics are
identical \cite{KACZER09,REISINGER10}. In particular, for a
measurement delay of \U{100}{\mu s}, a power-law in time
($\tstress^n$) with exponent $1/6$ is observed during stress while the
averaged recovery is roughly logarithmic over the relaxation time
$\trelax$.  This demonstrates that by using nanoscale devices, the
complex phenomenon of NBTI can be broken down to its microscopic
constituents: the defects that cause the discrete steps in the
recovery traces. Analysis of the statistics of these steps will thus
reveal the underlying physical principles.

\begin{figure}[!t]
\begin{center}
    \includegraphics[width=\figwidth,angle=-90]{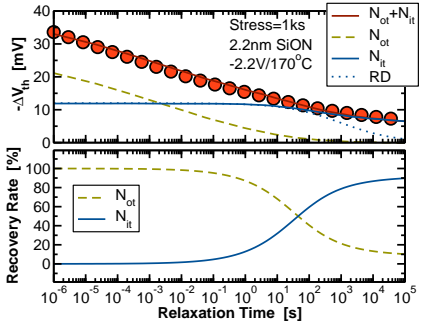}
    \vspace{-8mm}

    \capspace\caption{\label{f:relaxRD} \TFig: Recovery data (symbols) from our technology \cite{GRASSER11F} after
      \U{1}{ks} fitted by a simple hole trapping model ($\Not$) and the
      empirically modified RD model ($\Nit$), as taken from \cite{DESAI13}.  The dotted line (RD) shows the prediction of the unmodified RD model. \BFig: After about \U{50}{s}, according to this fit, 
      recovery is dominated by \emph{reaction-}limited $\Nit$ recovery. The recovery rate $R$ is defined by how much $\dVth$ is lost per decade in percent.
      \vspace*{0\baselineskip}
    }
\end{center}
\figspace
\end{figure}

It has been shown that the hole trapping component depends sensitively
on the process details, particularly for high nitrogen contents
\cite{MAHAPATRA13}, possibly making the choice
of benchmark technology crucial for our following arguments.  However,
for industrial grade devices with low nitrogen content such as those used in this study, no
significant differences in reported $\dVth$ drifts to published data have been found
\cite{REISINGER10}.  The pMOS samples used here are from a standard \U{120}{nm} CMOS
process with a moderate oxide thickness of \U{22}{\AA} and with a nitride content of approximately
6\%, while the poly-Si gates are boron doped with a
thickness of \U{150}{nm}.  In particular, our previously published
data obtained on the same technology as that of \Fig{f:SmallLarge} has
recently been interpreted
from the RD perspective \cite{DESAI13} as shown
in \Fig{f:relaxRD}, without showing any anomalies.  
This fit seems to suggest that after $\tstress =
\U{1}{ks}$ and $\trelax > \U{50}{s}$ recovery is dominated by \emph{diffusion}-limited $\Nit$
recovery, a conclusion we will put to the test in the following.

\section{Experimental Method}

For our experimental assessment we use the time-dependent defect
spectroscopy (TDDS) \cite{GRASSER10}, which has been extensively used
to study BTI in small-area devices at the single-defect level
\cite{WANG06,TOLEDANO11G,ZOU12B,GRASSER13}.  Since such devices
contain only a countable number of defects, the recovery of each
defect is visible as a discrete step in the recovery trace,
see \Fig{f:SmallLarge}.  The large variability of the discrete
  step-heights is a consequence of the inhomogeneous surface
potential caused by the random discrete dopants in the channel,
leading to percolation paths and a strong sensitivity of the
step-height to the spatial location of the trapped charge
\cite{ASENOV03B}. Typically, these step-heights are approximately
exponentially distributed \cite{KACZER10}
with the mean step-height given by $\aeta = \aetar \etaz$. Here,
$\etaz$ is the value expected from the simple charge sheet approximation
$\etaz = \q \tox / (\epsr \epsz W L)$,
where $\q$ is the elementary charge, $\epsr \epsz$ the permittivity of
the oxide, $W L$ the area, and $\tox$ the oxide thickness. Experiments
and theoretical values for the mean correction factor $\etar$ are in the
range $1$--$4$ \cite{FRANCO12}.

In a TDDS setup, a nanoscale device is repeatedly stressed and
recovered (say $N=100$ times) using fixed stress/recovery times,
$\tstress$ and $\trelax$.  
The recovery traces are analyzed for discrete steps of height $\eta$
occurring at time $\taue$. Each $(\taue, \eta)$ pair is then placed
into a 2D histogram, which we call the spectral map, formally denoted
by $g(\taue, \eta)$.  The clusters forming in the spectral maps reveal
the probability density distribution and thus provide detailed
information on the statistical nature of the average trap annealing
time constant $\ataue$.  From the evolution of $g(\taue, \eta)$ with
stress time, the average capture time $\atauc$ can be extracted as
well. So far, only exponential distributions have been observed for
$\taue$, consistent with simple independent first-order reactions
\cite{GRASSER12}.

In our previous TDDS studies, mostly short-term stresses ($\tstress
\lesssim \U{1}{s}$) had been used.  Based on this short-term nature,
the generality of these results may be questioned, since also $\Nit$
recovery predicted by RD models result in discrete steps
\cite{NAPHADE13B}.  As we have pointed out a while ago
\cite{GRASSER09C}, however, the distribution of these RD steps would
be loglogistic rather than exponential, a fact that should be clearly
visible in the spectral maps. In the following, we will conduct a
targeted search for such loglogistic distributions and other features
directly linked to \emph{diffusion}-limited recovery processes using extended
long-term TDDS experiments with $\tstress = \trelax = \U{1}{ks}$.

\section{Theoretical Predictions}

Before discussing the long-term TDDS data, we summarize the basic
theoretical predictions of the two model classes.  Both model classes
have in common that the charges trapped in interface and oxide states
induce a change of the threshold voltage.  Depending on the location
of the charge along the interface or in the oxide, it will contribute
a discrete step $\etai$ to the total $\dVth$. Due to only occasional
electrostatic interactions with other defects and measurement noise,
$\etai$ is typically normally distributed with mean $\aetai$.  The
mean values $\aetai$ themselves, however, are exponentially
distributed \cite{KACZER10}.  

The major difference between the model classes is whether creation and
annealing of $\Nit$ is \emph{diffusion-} or \emph{reaction}-limited, resulting in a
fundamentally different form of the spectral map $g(\taue, \eta)$, as
will be derived below.  Being the simpler case, we begin with the
dispersive \emph{reaction}-limited models.

\subsection{Dispersive Reaction-Limited Models}

In an agnostic formulation of dispersive \emph{reaction}-limited models,
creation and annealing of a single defect are assumed to be given by
a simple first-order reaction
\begin{align}
f(\tstress,\trelax,\atauc,\ataue) = \bigl(1-\exp(-\tstress/\atauc) \bigr) \exp(-\trelax/\ataue), \label{e:avg}
\end{align}
with $f$ being the probability of having a charged defect after stress
and recovery times $\tstress$ and $\trelax$, respectively.  The
physics of trap creation enter the average forward and backward time
constants $\atauc$ and $\ataue$.  It is important to highlight that equation \eq{e:avg} may describe both the
\emph{reaction}-limited creation and annealing of interface states
\cite{STESMANS96B,STESMANS00,HUARD06}, as well as a charge trapping
process \cite{HUARD06,WANG06,GRASSER10}.  We recall that even more
complicated charge trapping processes involving structural relaxation
and meta-stable defect states (such as switching oxide traps) can be approximately described by 
an effective first-order process, at least under quasi-DC conditions \cite{GRASSER12,GRASSER12C}.

Having $N$ defects present in a given device, the overall $\dVth$ is
then simply given by a sum of such first-order processes
\begin{align}
\dVth(\tstress, \trelax) = \sum_i \aetai f(\tstress,\trelax,\atauci,\atauei)  . \label{e:avgTrapping}
\end{align}
The most important aspect is that the time constants are observed to be widely
distributed. We have recently used such a model to explain BTI
degradation and recovery over a very wide experimental window assuming
the time constants to belong to two different distributions, one tentatively
assigned to charge-trapping and the other to interface state
generation \cite{KACZER09,GRASSER11F}. 

\begin{figure}[!t]
\begin{center}
    \includegraphics[width=\sfigwidth,angle=-0]{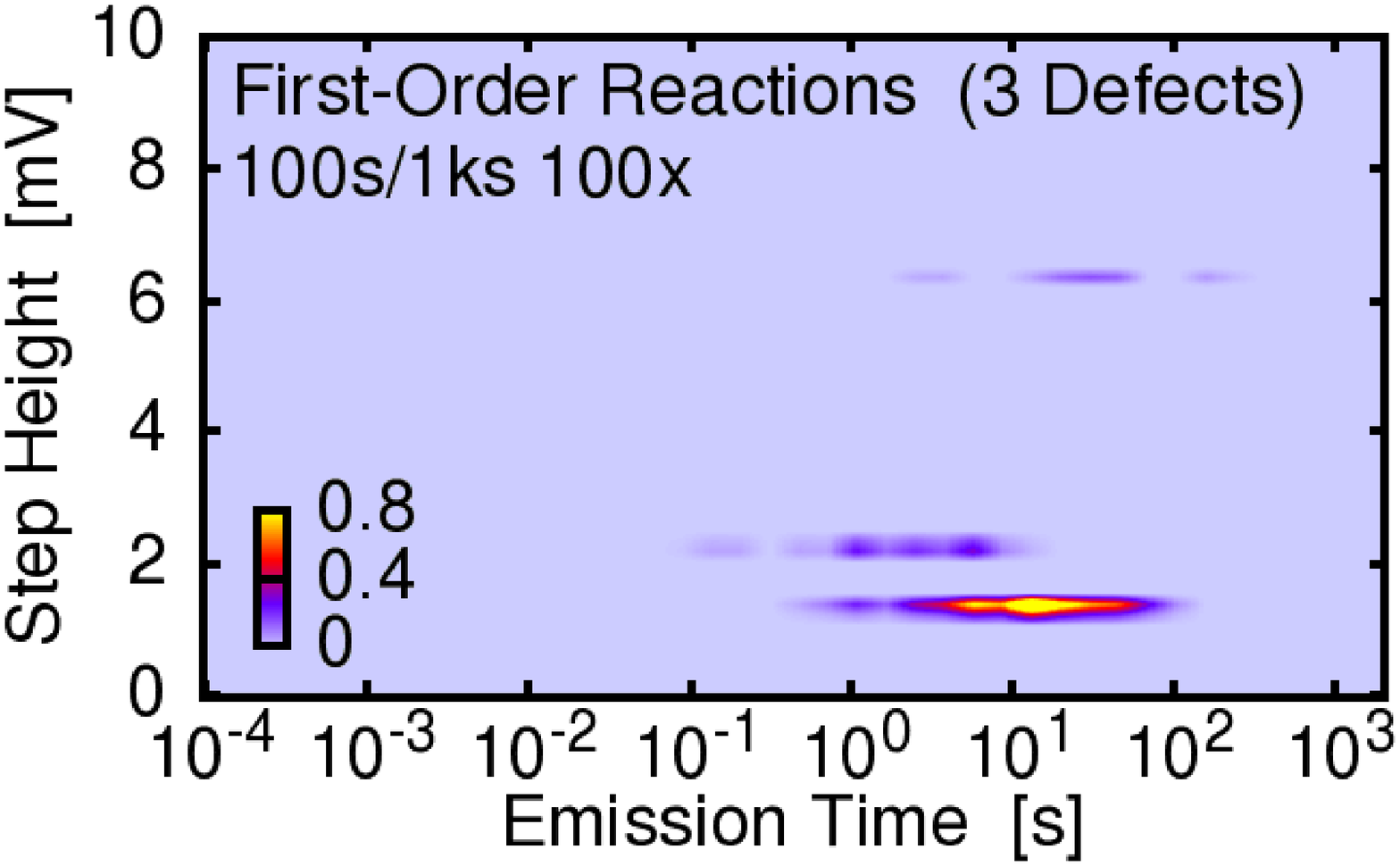}
    \includegraphics[width=\sfigwidth,angle=-0]{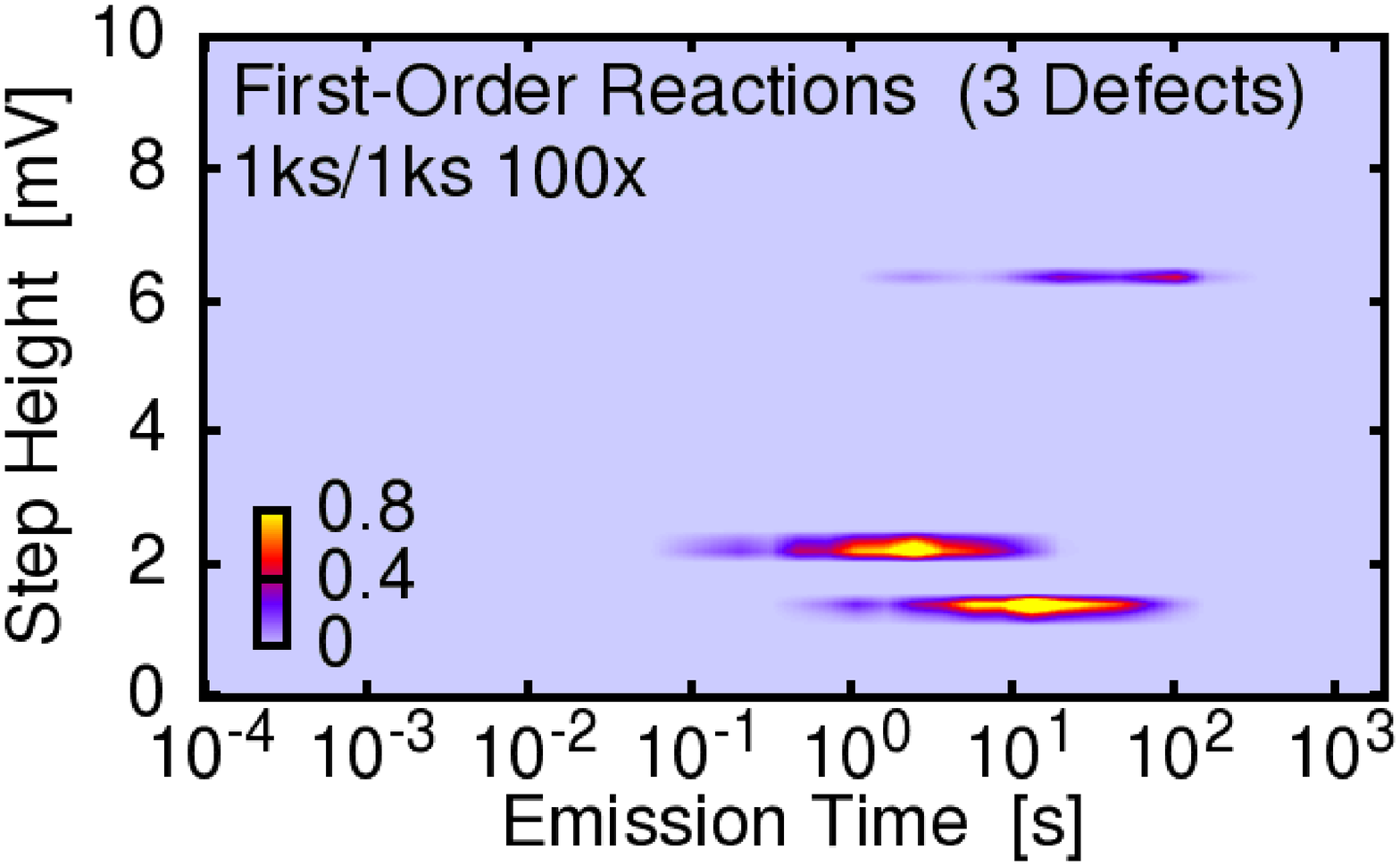}
    \capspace\caption{\label{f:NMP-ThreeDefects} Simulated spectral
      maps of a dispersive reaction model for three traps using two stress times, \U{100}{s} and \U{1}{ks} (left vs.~right).  The map is
      constructed using 100 repeated stress/relax cycles. The basic features
      are exponential clusters which do not move with stress time.
      \vspace*{0\baselineskip}
    }
\end{center}
\figspace
\end{figure}

At the statistical level, recovery in such a model is described by the
sum of exponential distributions. The spectral map, which records the
emission times on a logarithmic scale, is then given by
\begin{align}
g(\taue, \eta) = \sum_i B_i \, \feta\Bigl(\frac{\eta-\aetai}{\setai}\Bigr)\frac{\trelax}{\atauei} \exp(-\trelax/\atauei), 
\label{e:log-pdf}
\end{align}
with the stress time dependent amplitude $B_i \approx 1-\exp(-\tstress/\atauci)$ and $\feta$ describing the \pdf{}
of $\eta$, with mean $\aetai$ and standard deviation $\setai$.
An example spectral map simulated at two different stress times is
shown in \Fig{f:NMP-ThreeDefects}, which clearly reveals the three
contributing defects.
We note already here that contrary to the RD model, the spectral map
of the dispersive first-order model depends on the individual
$\atauei$, which can be strongly bias and temperature dependent.

\subsection{Non-Dispersive Reaction-Diffusion Models \label{s:RD}}

As a benchmark RD model we take the latest, and according to
  \cite{MAHAPATRA13} the physically most likely variant, the \pRD{}
  model: here it is assumed that \Hyd{} is released from
\Si\bondone\Hyd{} bonds at the interface, diffuses to the oxide-poly
interface, where additional \Si\bondone\Hyd{} bonds are broken to
eventually create \Hydt, the \emph{diffusion} of which results in the $n
=1/6$ degradation behavior typically associated with RD
models. Recovery then occurs via reversed pathways.  While other
  variants of the RD model have been used
  \cite{ALAM03,CHAKRAVARTHI04,ALAM07_MR,CHOI12B}, which
  cannot possibly be exhaustively studied here, we believe our
  findings are of general validity, as all these models are built around
\emph{diffusion}-limited processes.

In large-area devices the predicted long-term recovery after
long-term stress can be fitted by
 the empirical relation
\begin{align}
\Nit(\tstress, \trelax) \approx \frac{A \tstress^n}{1+ (\trelax/\tstress)^{1/s}} , \label{e:RD-relax}
\end{align}
with 
$s \approx 2$, provided diffusion is allowed into a semi-infinite gate stack with constant diffusivity in order to
avoid saturation effects. Quite intriguingly, a similar mathematical form has been
successfully used to fit a wide range of experimental data, using a
scaled stress time, though \cite{GRASSER07}.
Remarkably,
experimentally observed exponents
are considerably smaller than
what is predicted by RD models, corresponding to a wider
spread over the time axis.

In an empirically modified model, it has been assumed that in a real
3D device, recovery will take longer compared to \eq{e:RD-relax} since
the \Hyd{} atoms will have to ``hover'' until they can find a suitable
dangling bond for passivation \cite{MAHAPATRA13}.  However, using
a rigorous stochastic implementation of the RD model, we have
not been able to observe significant deviations from \eq{e:RD-relax},
irrespective of whether the model is solved in 1D, 2D, or 3D, provided
one is in the diffusion-limited regime \cite{SCHANOVSKY12}. As such,
significant deviations from the basic recovery behavior \eq{e:RD-relax} still
have to be rigorously justified.  One option to stretch the duration of recovery would be the consideration
of dispersive transport \cite{KACZER05B,ZAFAR05}.  Our attempts in this direction
were, however, not found to be in agreement with experimental observations \cite{GRASSER07,GRASSER11B}. 
Alternatively, consistent with experiment \cite{STESMANS00}, a distribution in the forward and backward reactions can be
introduced into the model \cite{CHOI12B}. This dispersion will stretch the distribution \eq{e:RD-relax}, i.e. increase the parameter $s$, 
but may also lead to a temperature dependence of the power-law slope, features which have not been validated
so far.  Nevertheless, a dispersion in the reaction-rates as used for instance in \cite{CHOI12B} will not
change the basic \emph{diffusion}-limited nature of the microscopic prediction as shown below.

\begin{figure}[!t]
\begin{center}
  \hbox{
    \includegraphics[width=\sfigwidth,angle=-90]{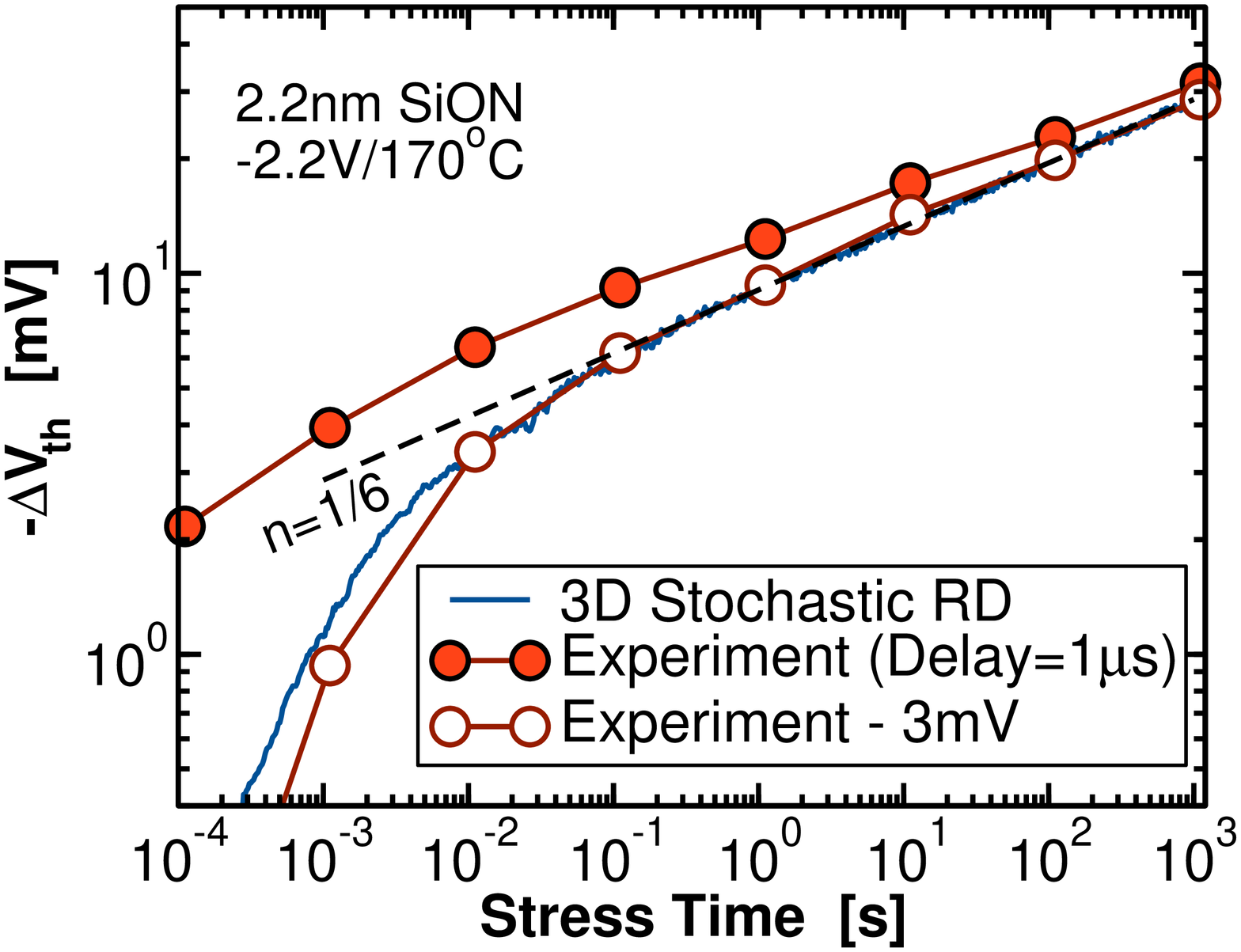}
    \hspace{-4mm}

    \includegraphics[width=\sfigwidth,angle=-90]{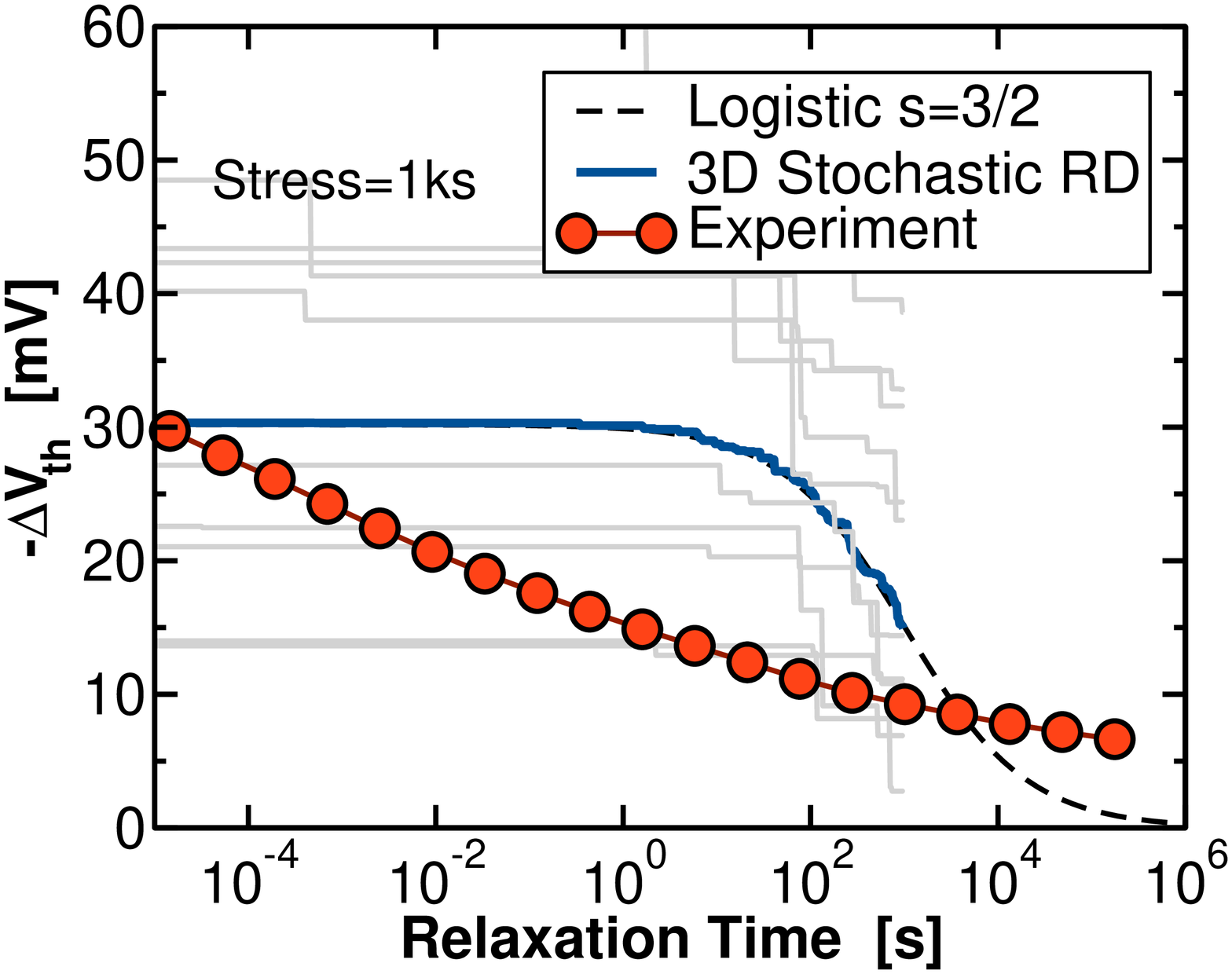}
  }
    \vspace{-6mm}

    \capspace\caption{\label{f:RD-check} Degradation (left) and recovery (right) predicted by the calibrated stochastic \pRD{} model
      on small-area devices.
      The difference in the initial stress phase is assumed to be due to hole trapping and approximately modeled
      by subtracting \U{3}{mV} from the experimental data,
      as we are here concerned with larger stress and recovery times, where hole trapping is assumed to be negligible
      in the RD interpretation \cite{MAHAPATRA09B}.  Recall that
      \Fig{f:relaxRD} uses the \emph{empirically stretched} RD model \cite{DESAI13}.
      \vspace*{-1\baselineskip}
    }
\end{center}
\figspace
\end{figure}

In order to study the stochastic response of the \pRD{} model, we extended our previous stochastic
implementation \cite{SCHANOVSKY12} of the \Hyd/\Hydt{} RD model to include the oxide/poly
interface following ideas and parameters of \cite{NAPHADE13}.  Since any sensible macroscopic model
is built around a well-defined microscopic picture, in this case non-dispersive diffusion and non-dispersive
rates, these features of the microscopic model must be preserved in the macroscopic theory, leaving little
room for interpretation. In order to be
consistent with the $W \times L=\U{150}{nm} \times \U{100}{nm}$
devices used in our TDDS study, we chose $\aeta = \aetar \etaz = 2 \times \U{0.9}{mV}
= \U{1.8}{mV}$ \cite{GRASSER10}.
Furthermore, a typical density
of interface states $\Nit = 2 \times \U{10^{12}}{cm^{-2}}$
\cite{STESMANS00,CHOI12B} is assumed. We would thus expect
about 300 such interface states to be present for our TDDS devices. %

Before looking into the predictions of this RD model in a TDDS
setting, we calibrate
our implementation of the \pRD{} model to experimental stress
data, see \Fig{f:RD-check} (left).  
In order to obtain a good fit, we follow the procedure suggested in \cite{MAHAPATRA09B} and subtract a virtual hole trapping contribution of \U{3}{mV} from the experimental data to obtain the required $n=1/6$ power-law.
Also, we remark that to achieve this fit, unphysically
large hydrogen hopping distances had to be used in the microscopic model \cite{SCHANOVSKY12}. Furthermore,
\Hydt{} had to be allowed to diffuse more than a micrometer deep into
the gate stack with unmodified diffusion constant to maintain the $n=1/6$ power-law
exponent, despite the
fact that our poly-Si gate was only \U{150}{nm} thick.

From \eq{e:RD-relax} we can directly calculate the expected
unnormalized probability density function for RD recovery
as
\begin{align}
\fRD(\trelax) = - \PD{\Nit(\trelax)}{\log(\trelax)} = A \tstress^n \, \frac{(\trelax/\tstress)^{1/s}}{s \bigl(1+ (\trelax/\tstress)^{1/s}\bigr)^2}  \label{e:RD-relax-pdf-single}
\end{align}
which after normalization by $A \tstress^n$ is a loglogistic
distribution of $\log(\trelax)$ with parameter $s$ and mean
$\log(\tstress)$.  In the framework of the standard
  non-dispersive RD model, all interface states are equivalent in the
sense that on average they will have degraded and recovered with the
same probability at a certain stress/recovery time combination.  In
terms of impact on $\dVth$, we again assume that the mean impact of a
single trap $\aetai$ is exponentially distributed.

\begin{figure}[!t]
\begin{center}
    \includegraphics[width=\mapwidth,angle=-0]{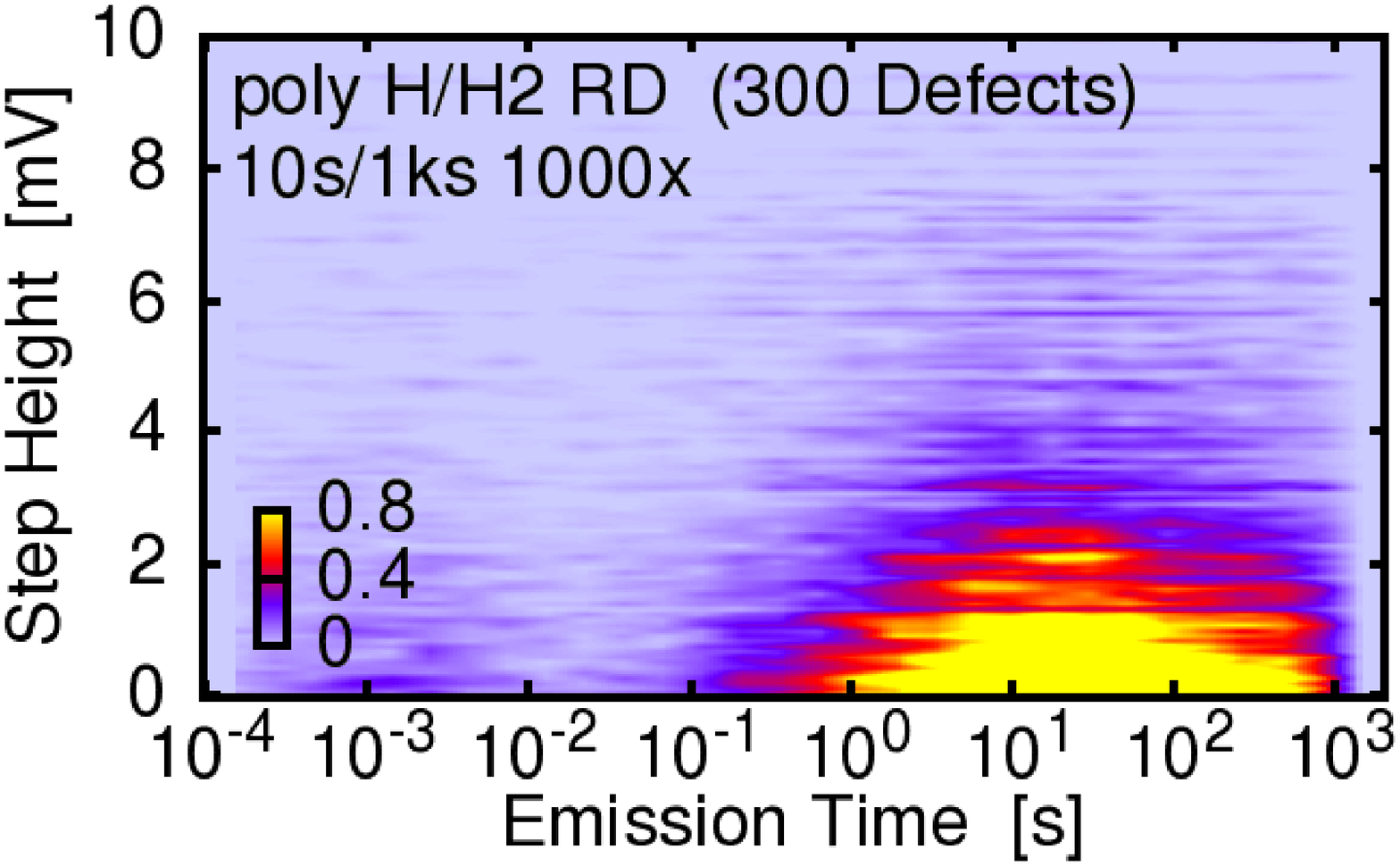}
    \includegraphics[width=\mapwidth,angle=-0]{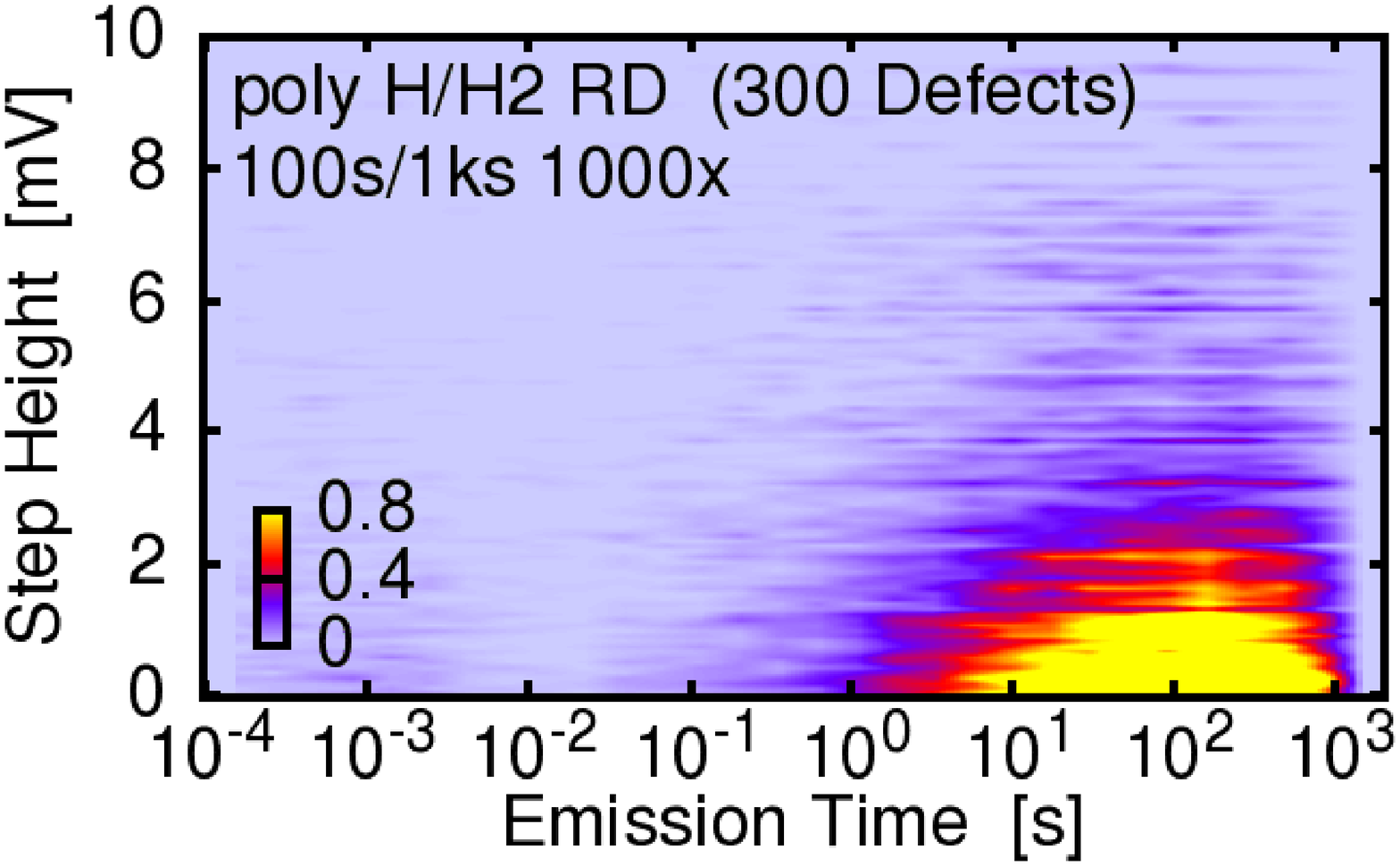}
    \capspace\caption{\label{f:RD-NumberOfDefects} Since in non-dispersive RD models \emph{all} defects contribute equally to the spectral map,
      no clear clusters can be identified, except for possibly in the tail of the exponential distribution. Shown is a \pRD{} simulation
      with 300 defects for two stress times. Note that on average all defects are active with the same probability at all times, which results in markedly different spectral
      maps compared to those produced by a dispersive reaction model (\Fig{f:NMP-ThreeDefects}).
      \vspace*{0\baselineskip}
    }
\end{center}
\figspace
\end{figure}

Using \eq{e:RD-relax-pdf-single}, the spectral map built of subsequent stress/relax cycles can
be obtained.
Since except for their step-heights all defects are equivalent, the time dynamics can be pulled out
of the sum to eventually give
\begin{align}
g(\taue, \eta) = A \tstress^n \, \frac{(\trelax/\tstress)^{1/s}}{s \bigl(1+ (\trelax/\tstress)^{1/s}\bigr)^2} \sum_i \feta\Bigl(\frac{\eta-\aetai}{\setai}\Bigr)
   \label{e:RD-relax-pdf} .
\end{align}
This is a very interesting result, as it implies that all defects are
active with the same probability at any time, leading to a dense response in the
spectral map as shown in \Fig{f:RD-NumberOfDefects}.  As will be
shown, this is incompatible with our experimental results.

\begin{figure}[!t]
\begin{center}
    \includegraphics[width=\mapwidth,angle=-0]{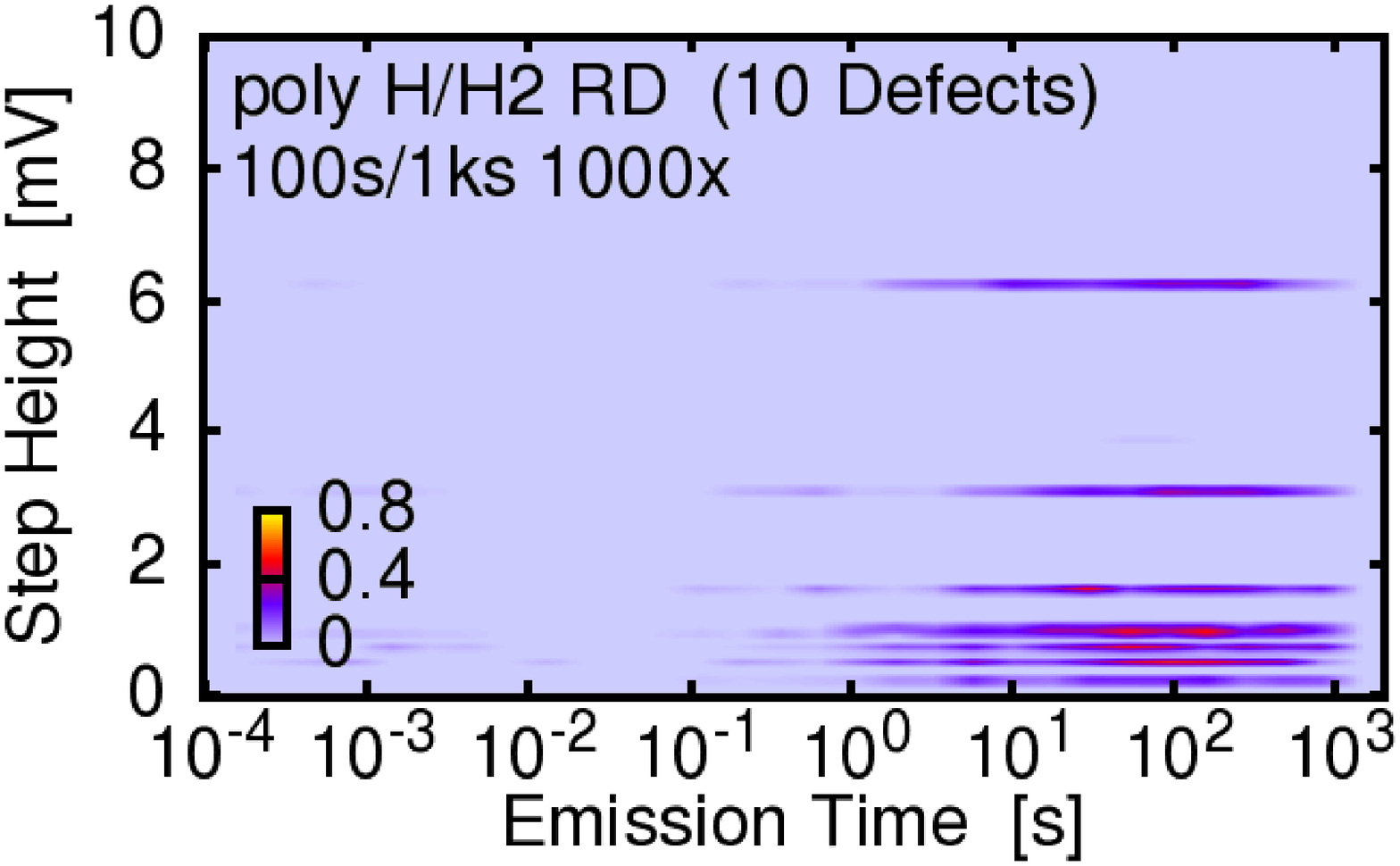}
    \includegraphics[width=\mapwidth,angle=-0]{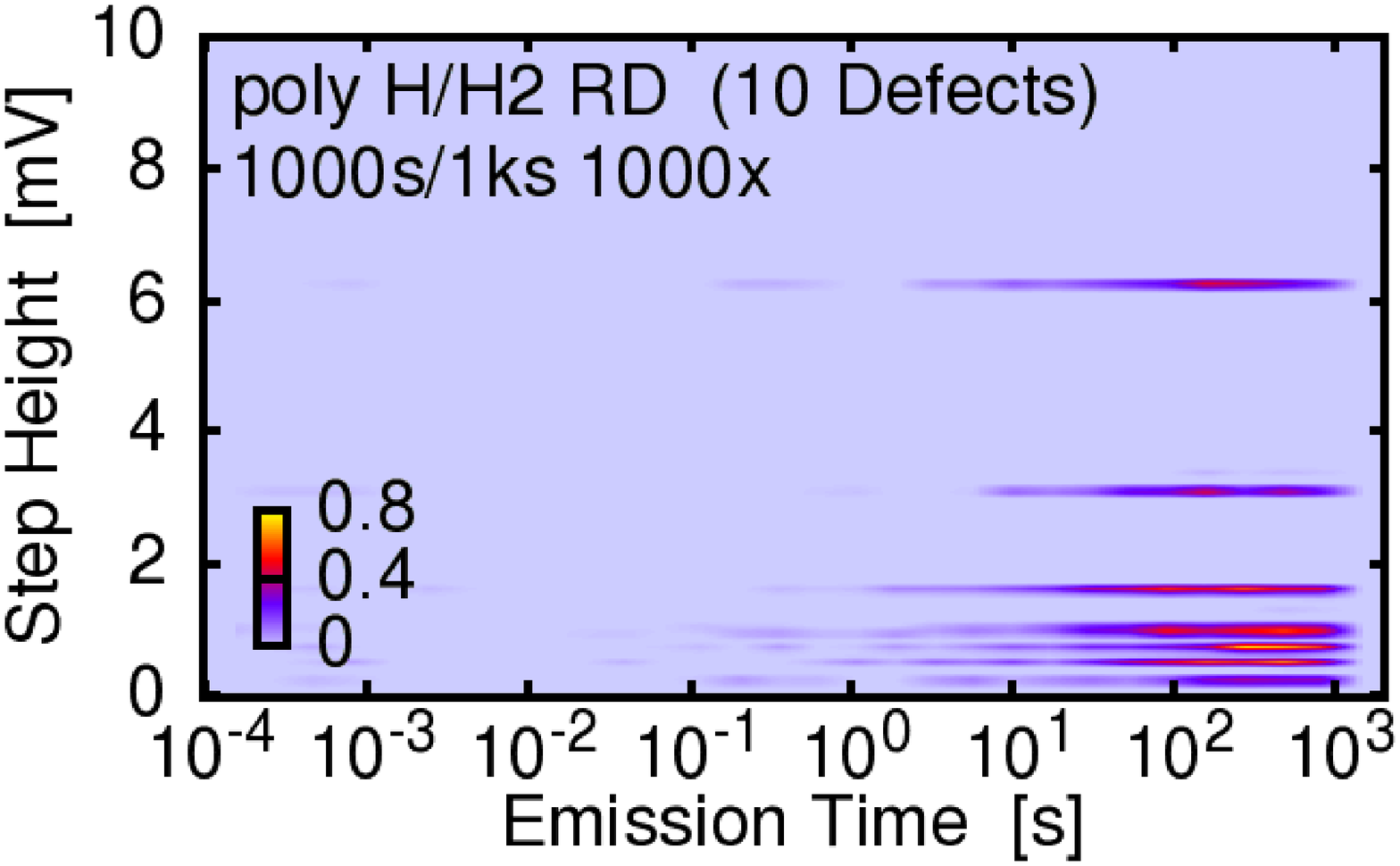}

    \capspace\caption{\label{f:RD-SM} Simulated spectral maps using the \pRD{} model for a $\U{20}{nm} \times \U{25}{nm}$ sized device with only
      ten active defects.  Note how the intensity, that is the emission probability, of the clusters keeps increasing while the mean of the clusters shifts to
      larger times with increasing stress time.
    }
\end{center}
\vspace*{-5mm}
\end{figure}

In order to more clearly elucidate the features of the RD
model, we will in the following use a $\U{20}{nm} \times \U{25}{nm}$
device, in which only a small number of defects (about ten) contribute to the
spectral maps.
The crucial fingerprint of the RD model would then be that these clusters
are loglogistically distributed and thus much wider than the previously
observed exponential distributions.
Furthermore, we note that the RD spectral map does not depend on any
parameter of the model nor does it depend on temperature and bias
{\cite{ALAM03}, but due to the \emph{diffusion}-limited nature of the model}
shifts to larger times with increasing stress time (see
\Fig{f:RD-SM}),
facts we will compare against experimental data
later.

\section{Small-Devices: Purely Reaction-Limited}

As noted before, previous TDDS experiments had been limited to stress
times mostly smaller than about \U{1}{s}, which may limit the
relevance of our findings for long-term stress.  As such, it was
essential to extend the stress and relaxation times to \U{1}{ks},
which is a typically used experimental window \cite{MAHAPATRA13}.
Unfortunately, the stress/relax cycles needed to be repeated at least
100 times, otherwise differentiation between exponential and logistic
distributions would be difficult.  We therefore used 9 different
stress times for each experiment, starting from \U{10}{\mu s} up to
\U{1}{ks} with recovery lasting \U{1}{ks}, repeated 100 times,
requiring a total of about 12 days.  About 20 such experiments were
carried out on four different devices over the course of more than
half a year.

\begin{figure}[!t]
\begin{center}
      \includegraphics[width=\mapwidth,angle=0]{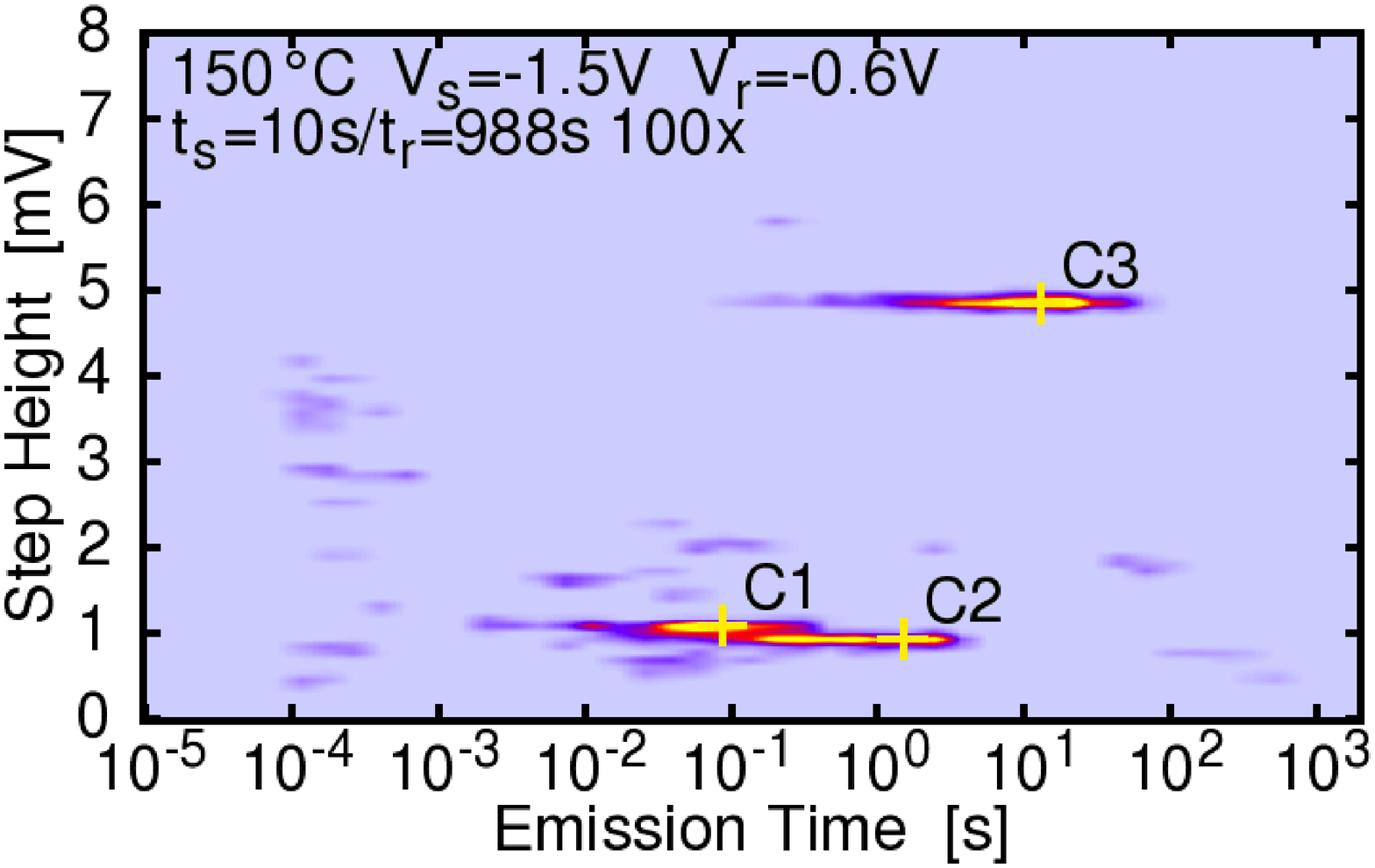}
      \includegraphics[width=\mapwidth,angle=0]{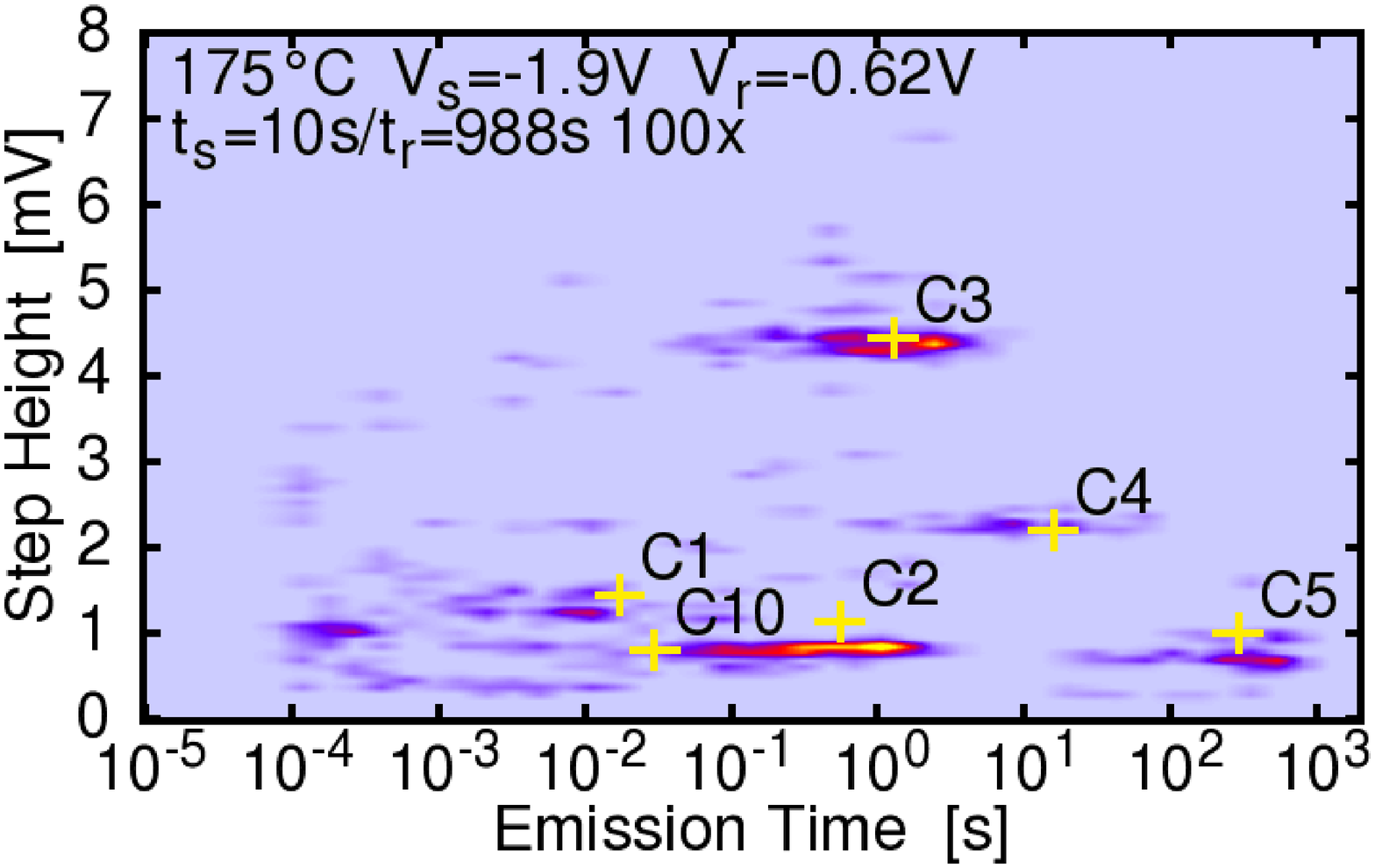}

      \includegraphics[width=\mapwidth,angle=0]{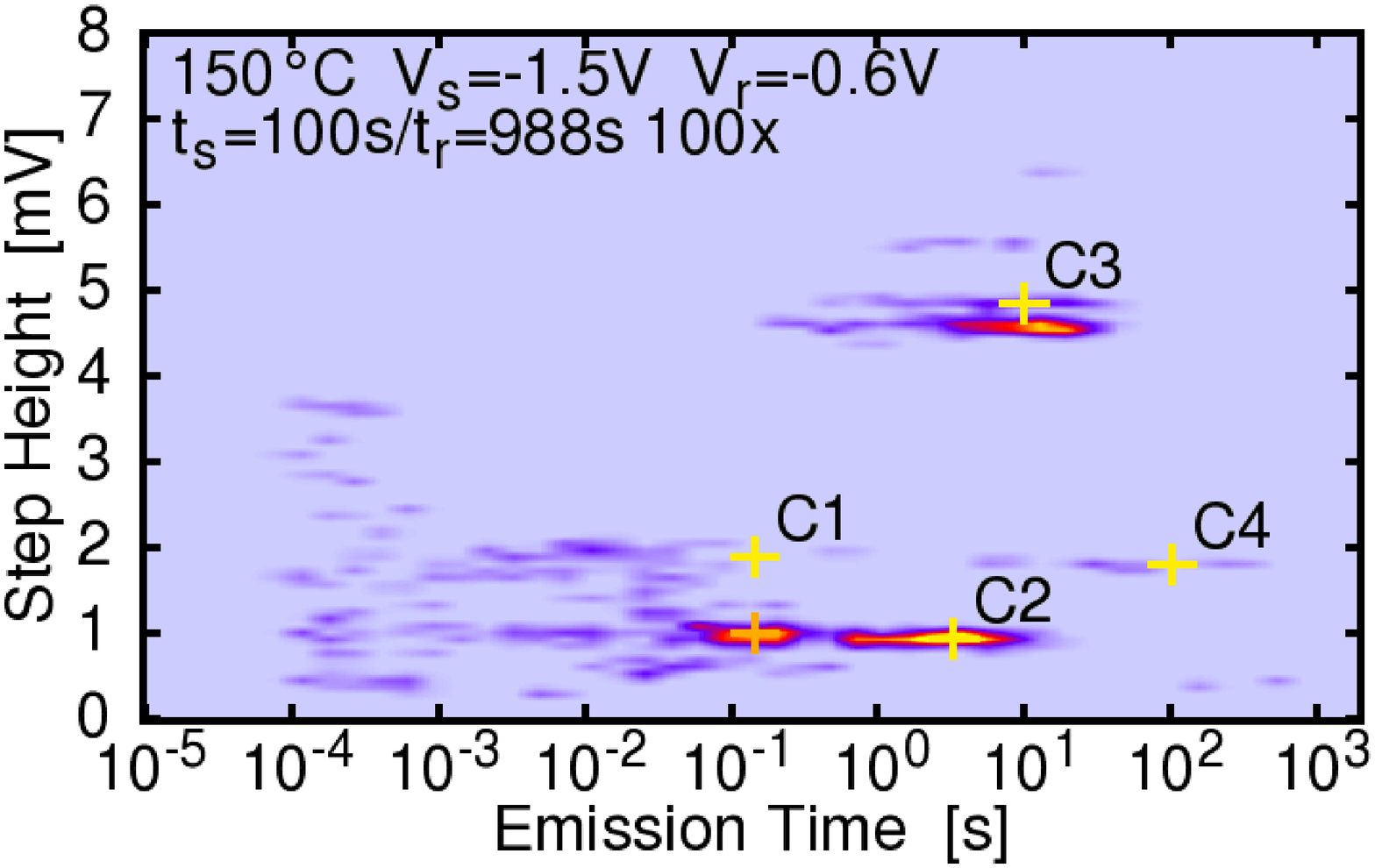}
      \includegraphics[width=\mapwidth,angle=0]{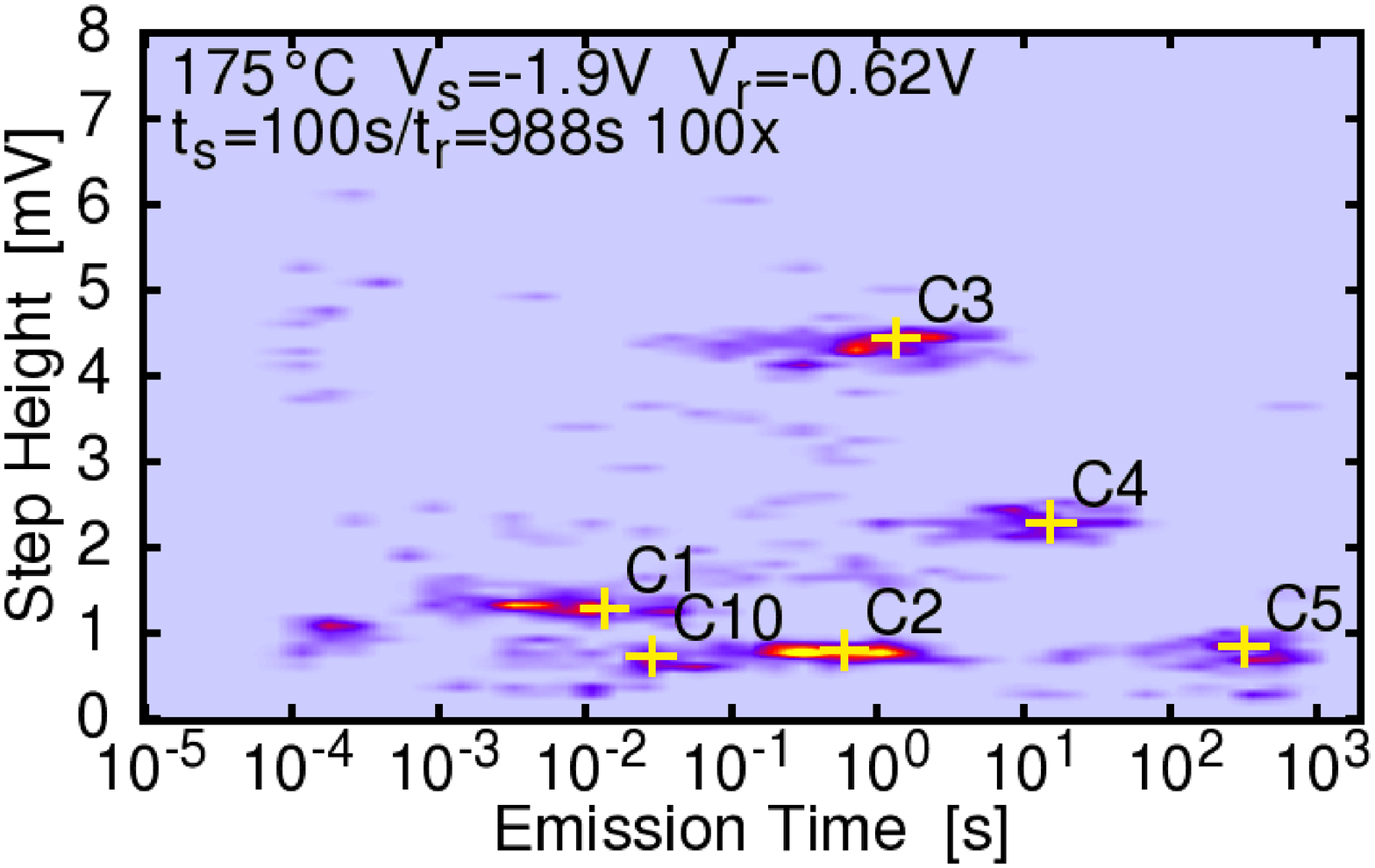}

      \includegraphics[width=\mapwidth,angle=0]{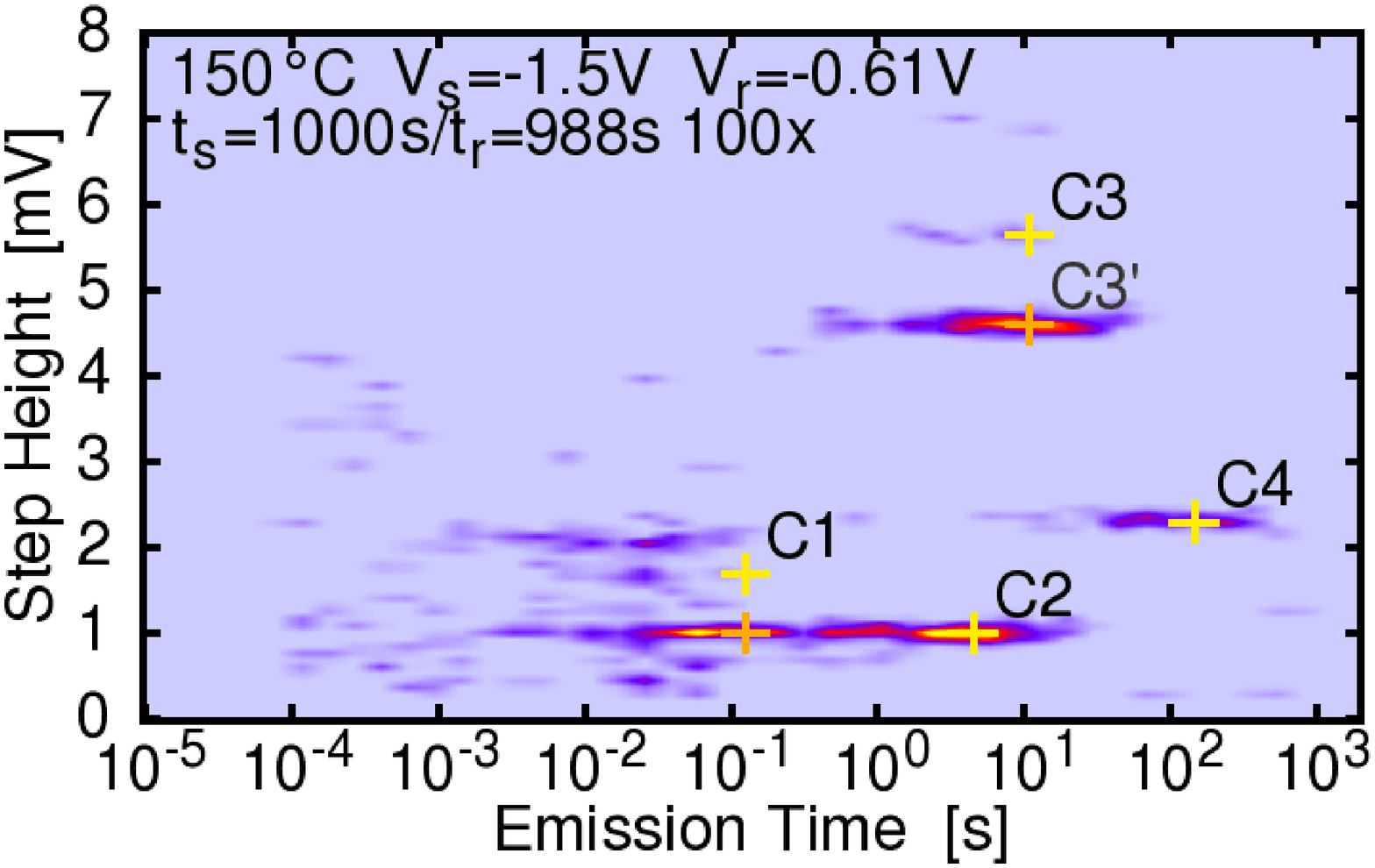}
      \includegraphics[width=\mapwidth,angle=0]{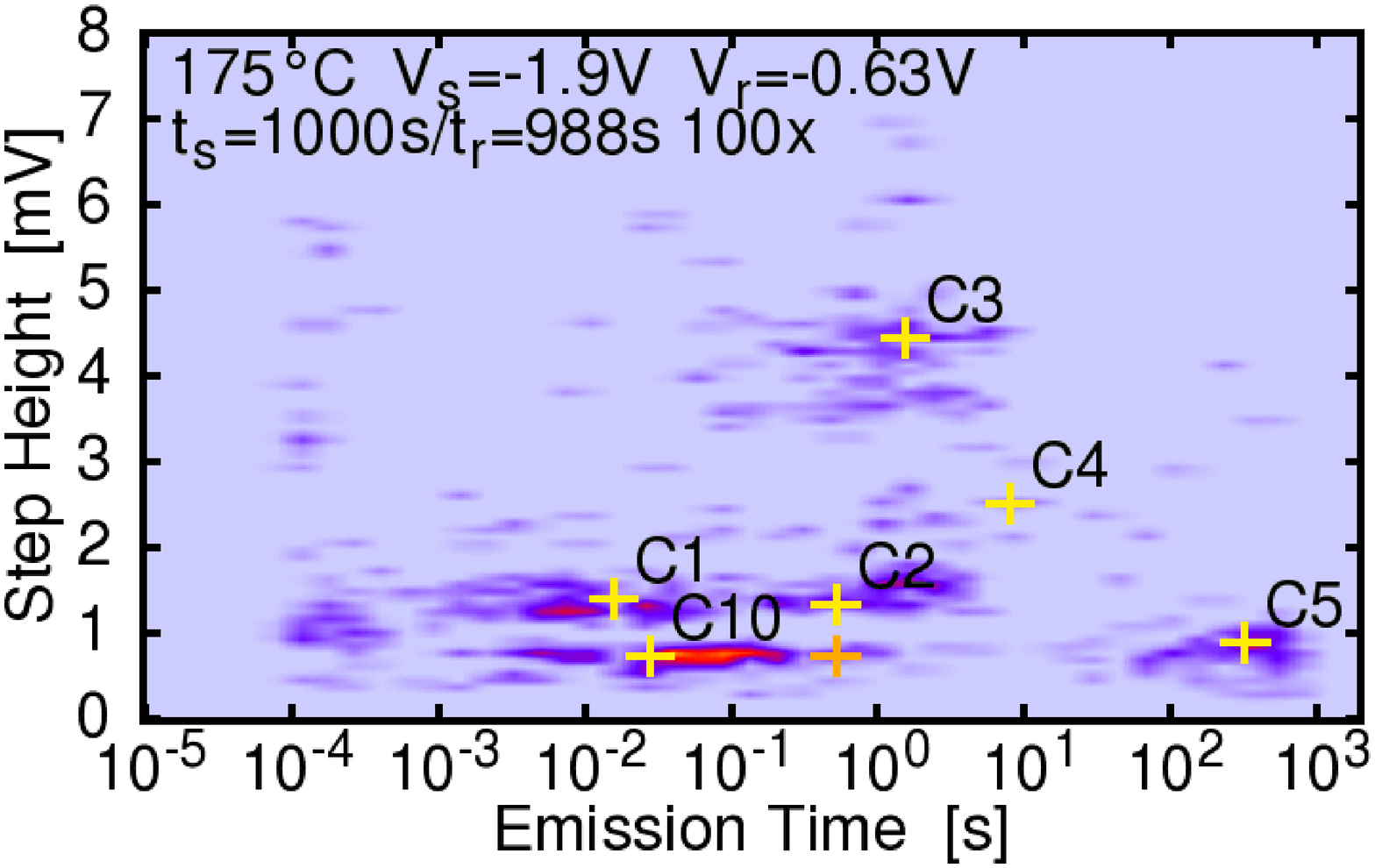}

      \includegraphics[width=3.9cm,angle=-90]{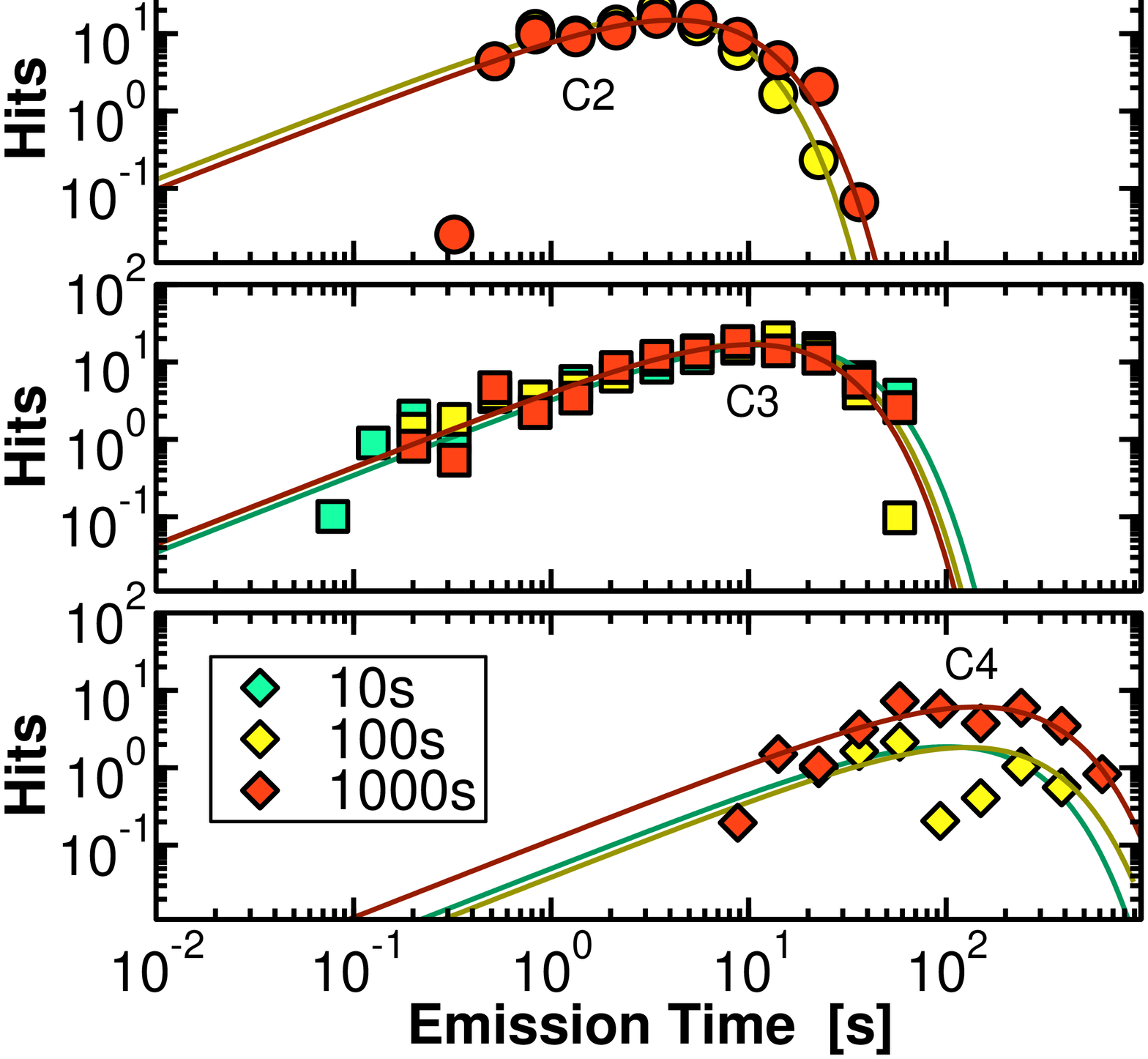}
      \includegraphics[width=3.9cm,angle=-90]{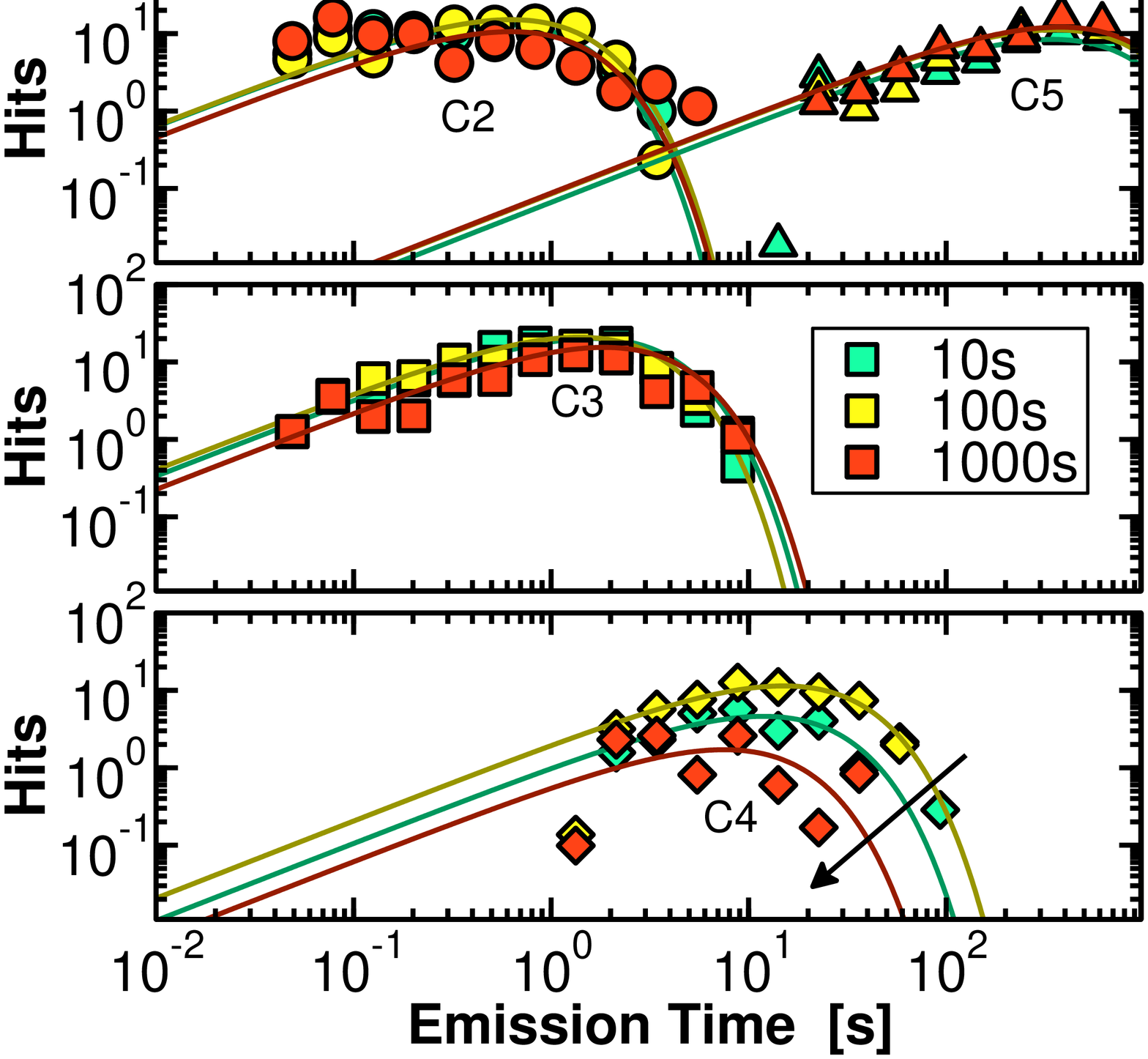}

    \caption{\label{f:LongTermTDDS-Maps} Even at longer stress times
      (\U{10}{s} -- \U{1}{ks}) and higher temperatures, 150\degC{} (left/top) and 175\degC{} (right/top),
        all clusters (symbols) are exponential (lines) and do not move with stress time (bottom),
        just like the prediction of a \emph{reaction}-limited model, see \Fig{f:NMP-ThreeDefects}. Due to the increasing number
      of defects contributing to the emission events, the data
      becomes noisier with increasing stress bias, temperature, and
      time. With increasing stress, defect C4 shows signs
        of volatility, leading to a smaller number of emission events at longer times \cite{GRASSER13B}.
}
\end{center}
\figspace
\end{figure}

Since we are particularly interested in identifying a
\emph{diffusion}-limited contribution to NBTI recovery, we tried to
minimize the contribution of charge trapping.  With increasing stress
voltage, an increasing fraction of the bandgap becomes accessible for
charging \cite{GRASSER12}, which is why we primarily used stress
voltages close to $\VDD = \U{-1.5}{V}$ of our technology (about
\U{4}{MV/cm} \cite{REISINGER08}).  Furthermore, it has been observed
that at higher stress voltages defect generation in a TDDB-like
degradation mode can become important
\cite{MAHAPATRA04,MAHAPATRA11,MAHAPATRA13}, an issue we avoid at such
low stress voltages.
Two example
measurements are shown in \Fig{f:LongTermTDDS-Maps} for \U{-1.5}{V} at
150\degC{} and \U{-1.9}{V} at 175\degC{} (about 4 -- \U{5}{MV/cm}).
As already observed for short-term stresses, all
clusters are exponential and have a temperature-\emph{dependent} but
time-\emph{independent} mean $\ataue$.  Most noteworthy is the fact that \emph{no sign of an RD
  signature as discussed in \Sec{s:RD} was observed}.  We remark that defects
tend to show strong signs of volatility at longer stress and recovery
times \cite{GRASSER13B}, a fascinating issue to be discussed in more
detail elsewhere.

\begin{figure}[!t]
\begin{center}
    \includegraphics[width=5.8cm,angle=0]{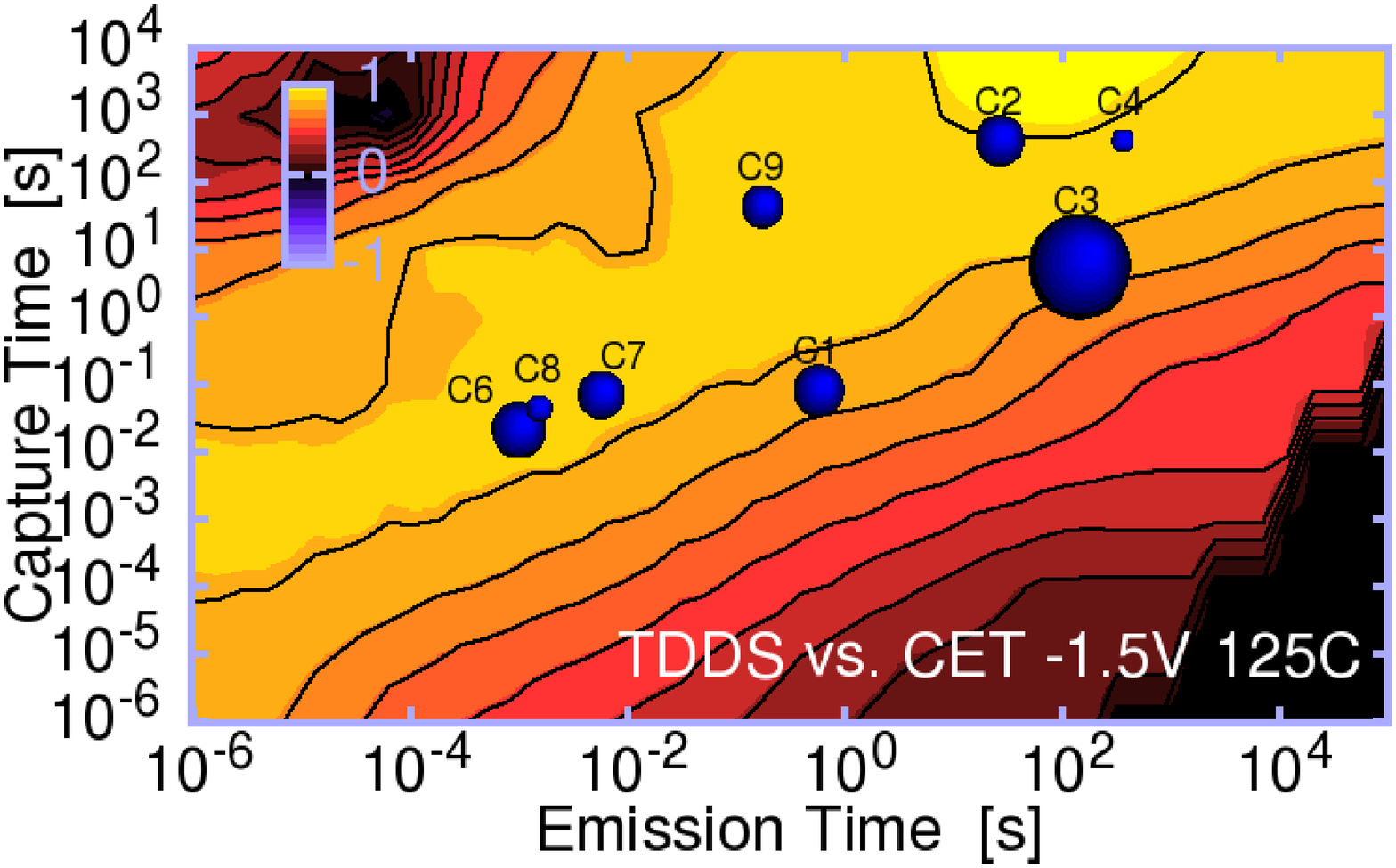}
    \vspace*{-5mm}

    \includegraphics[width=\figwidth,angle=-90]{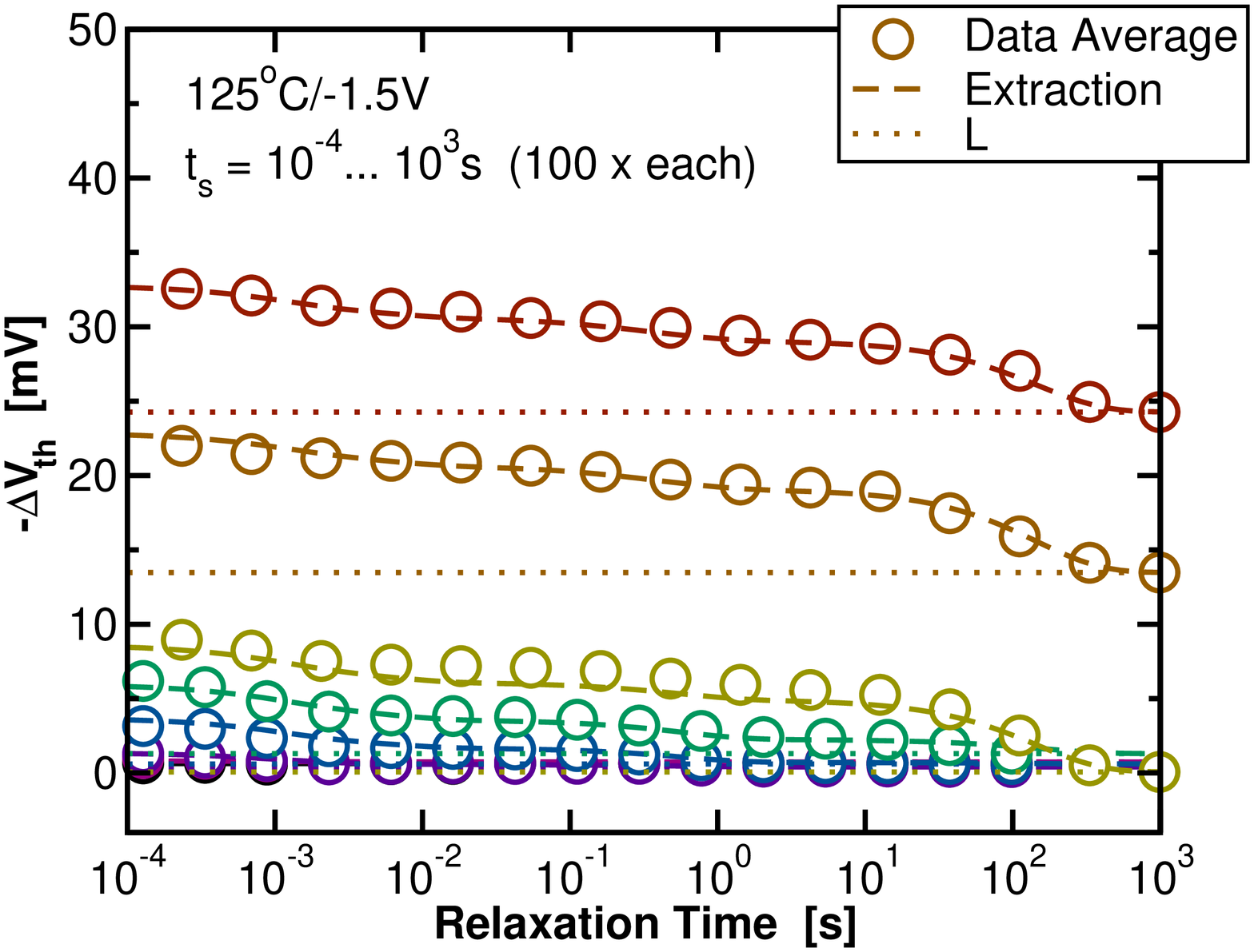}
    \vspace{-10mm}

    \capspace\caption{\label{f:LongTermTDDS-Fit} 
      \TFig: Comparison of the extracted capture and emission times vs. a capture/emission time
      (CET) map from a large area device \cite{GRASSER11F}.  The size of the dots represents the $\eta$ value
      of each defect.  The distribution of the individual defects seen in TDDS
      agrees well with the CET map.
      \BFig: Using the time constants extracted from the long-term TDDS data (lines),
      it is possible to \emph{fully} reconstruct the average recovery traces (symbols, corresponding to the expectation
      value) for all stress and recovery times.  The average offset $L$ at $\trelax = \U{1}{ks}$ is
      added (dotted lines) to show the build-up of defects with larger emission/annealing times ($\sim$ permanent component).
    }
\end{center}
\figspace
\end{figure}

To confirm that our extracted capture and emission times fully
describe recovery \emph{on average}, we calculate the average of all 100
recovery traces recorded at each stress time and compare it with the
prediction given by the extracted $\atauci$ and $\atauei$ values using
\eq{e:avgTrapping}, which corresponds to the expectation value and
thus the average.  Indeed, as shown in \Fig{f:LongTermTDDS-Fit},
excellent agreement is obtained, finally proving that our extraction
captures the essence of NBTI recovery.  It is worth
  pointing out that this agreement is obtained \emph{without} fitting of the
  average data: we simply use the extracted capture and emission times as
well as the extracted step-heights and put them into \eq{e:avgTrapping}.  Also shown is
  a comparison of the capture/emission times extracted by TDDS with a
  capture/emission time (CET) map extracted on large devices
  \cite{GRASSER11F}.  The capture and emission times extracted on
  the nanoscale device are fully consistent with the macroscopic
  distribution and correspond to a certain \emph{realization},
  which will vary from device to device.

\begin{figure}[!t]
\begin{center}
  \hbox{
    \includegraphics[width=\sfigwidth,angle=-90]{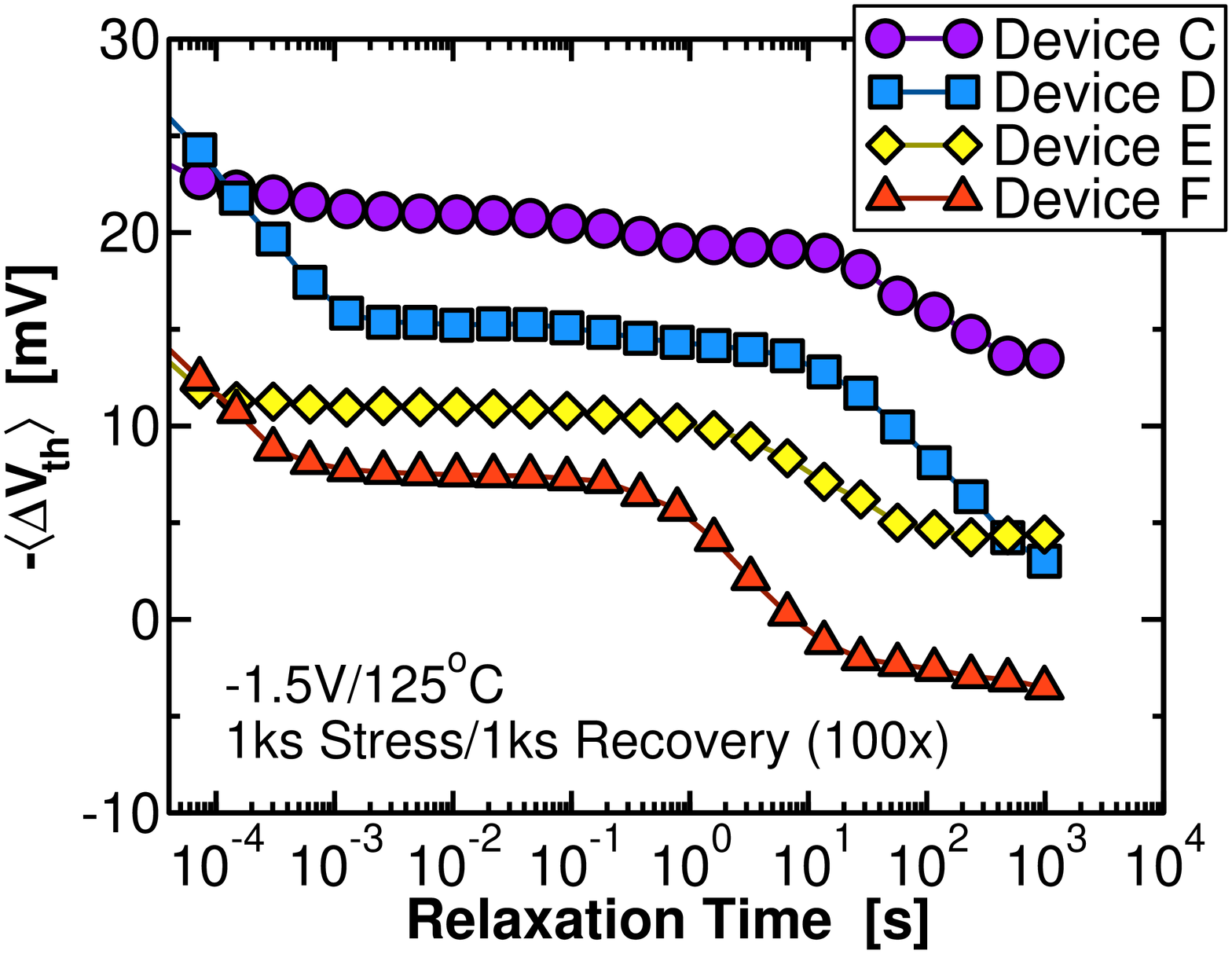}
    \hspace{-4mm}
    \includegraphics[width=\sfigwidth,angle=-90]{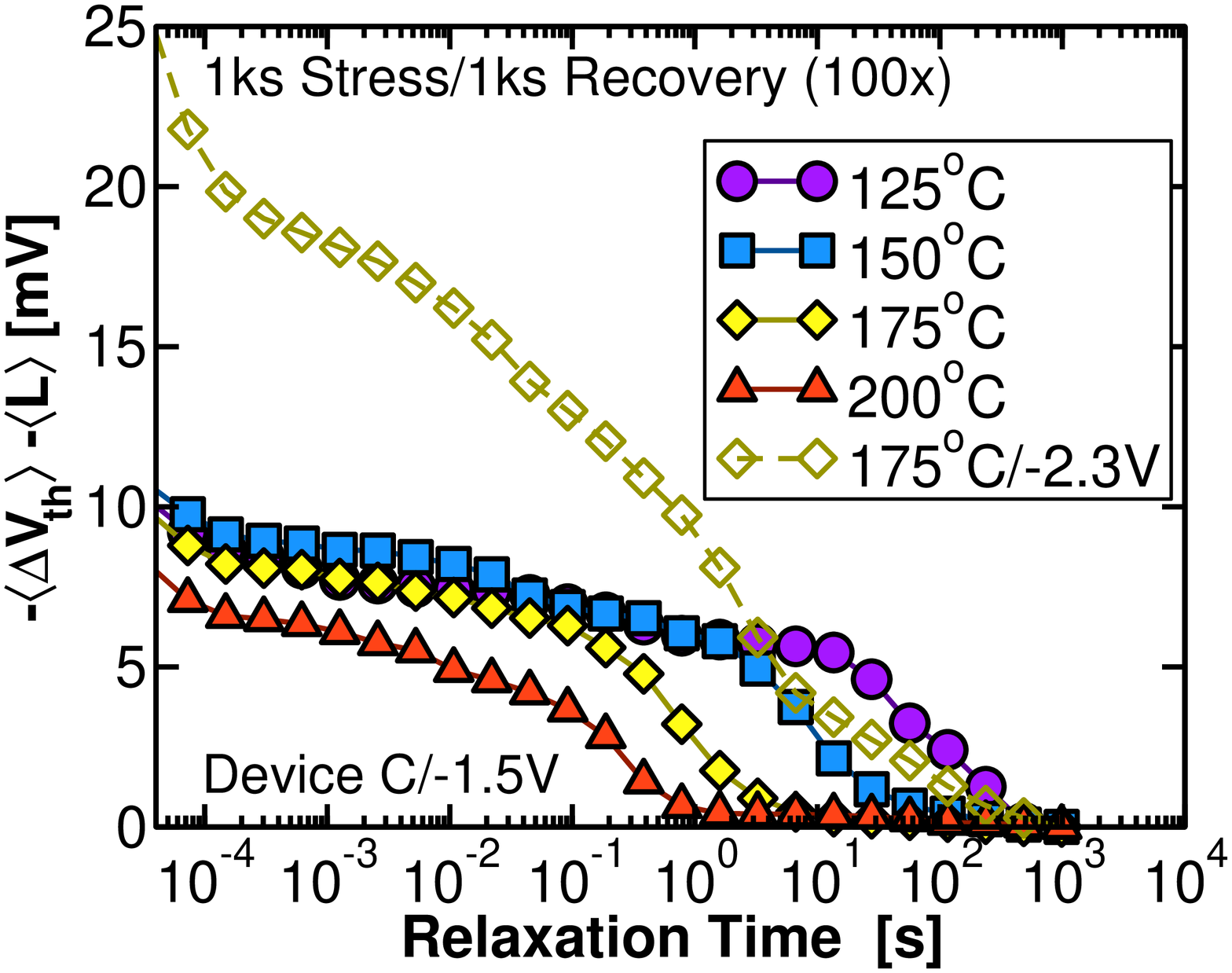}
  }
    \vspace{-6mm}

    \capspace\caption{\label{f:Dev2Dev} \LFig: Just like for short-term
      stress, the averaged recovery over 100 repetitions after long-term stress/relax cycles varies strongly from
      device to device.  Also noteworthy is the dramatic difference
      in the last value at $\trelax = \U{1}{ks}$ of each averaged recovery trace, meaning that
      also the build-up of the permanent component is stochastic.
      \RFig: Contrary to the RD prediction, recovery depends strongly on temperature and stress bias.
      Shown is the average recovery data minus the averaged last value of each recovery
      trace $\statav{L}$.
      In device C, due to the strong temperature activation of the emission
      time constants, the average $\dVth$ traces are shifted to shorter
      times with increasing $T$, leading to practically no recovery
      after \U{1}{s} for $\Vstress = \U{-1.5}{V}$. At higher
      stress voltages ($\Vstress = \U{-2.3}{V}$), a considerably
      larger number of traps (presumably in the oxide) can contribute, leading to
      strong recovery in the whole measurement window.
      \vspace*{0\baselineskip}
    }
\end{center}
\figspace
\end{figure}

As a final point, we compare the \emph{averaged} recovery over 100
repetitions obtained from four different nanoscale devices after
\U{1}{ks} stress under the same conditions in \Fig{f:Dev2Dev}.
Clearly, all devices recover in a very unique way.  For instance,
device F shows practically no recovery between \U{10}{s} and \U{1}{ks}
while device D has a very strong recoverable component in this time
window but practically no recovery from \U{1}{ms} up to \U{10}{s}.
Furthermore, this unique recovery depends strongly on bias and
temperature, as demonstrated in \Fig{f:Dev2Dev} (right) for device C.  For
example, after a stress at \U{-1.5}{V} at 125\degC{}, strong recovery
is observed between \U{10}{s} and \U{1}{ks}, which is completely
absent at 200\degC.  On the other hand, if the stress bias is
increased to say \U{-2.3}{V} (about \U{7}{MV/cm}), a nearly logarithmic recovery is
observed in the whole experimental window, consistent with what is
also seen in large-area devices.  

In the non-dispersive RD picture, hundreds of defects would be
  equally contributing to the average recovery of such devices.  As
  such, the model is practically immune to the spatial distribution of
  the defects which would be the dominant source of device-to-device
  variability in this non-dispersive RD picture, lacking any other
  significant parameters.  Such a model can therefore not explain the
  strong device-to-device variations observed experimentally.
Also, as discussed before, in non-dispersive RD models in their
present form recovery is independent of bias and temperature, which is
also at odds with these data.

On the other hand, our data is perfectly consistent with a collection
of defects with randomly distributed $\atauci$ and $\atauei$.  In this picture, the
occurrence of a recovery event only depends on whether a defect with a
suitable pair ($\atauei$, $\atauci$) exists in this particular
device.  Since these time constants depend on bias and temperature,
the behavior seen in \Fig{f:Dev2Dev} is a natural consequence.

\section{Consequences}

The question whether NBTI is due to a \emph{diffusion-} or
\emph{reaction-}limited process is of high practical significance
and not merely a mathematical modeling detail.  First of all, it is
essential from a process optimization point of view: if the RD model
in any variant were correct, then one should seek to prevent the
diffusion into the gate stack by, for instance, introducing hydrogen
diffusion barriers.  This is because according to RD models, upon
hitting such a barrier, the hydrogen concentration in the gate stack
would equilibrate, leading to an end of the degradation.  On the other
hand, if \emph{reaction-}limited models are correct -- and our results
clearly indicate that they are -- device optimization from a reliability
perspective should focus on the distribution of the time
constants/reaction rates in the close vicinity of the channel that are
responsible for charge trapping and the \emph{reaction-}limited
creation of interface states.  

Secondly, our results have a fundamental impact on our understanding
of the stochastic reliability of nanoscale devices.  We have
demonstrated that even the averaged response of individual devices
will be radically different from device to device, whereas in
non-dispersive RD models all devices will \emph{on average} degrade in
the same manner.  Given the strong bias- and temperature-dependence of
this individual response, it is mandatory to study and understand the
distribution of the bias- and temperature-dependence of the
responsible reaction-rates. This is exactly the route taken recently
in \cite{FRANCO13}, where it was shown that the energetic alignment of
the defects in the oxide with the channel can be tuned by modifying
the channel materials in order to optimize device reliability.  

\section{Conclusions} 
Using nanoscale devices, we have established \emph{an ultimate
    benchmark} for BTI models at the statistical level.  Contrary to
  previous studies, we have used a very wide experimental window,
  covering stress and recovery times from the microsecond regime up to
  kiloseconds, as well as temperatures up to 175\degC.  The crucial
  observations are the following:
\begin{itemize}
\item[\mycnt] Using time-dependent defect spectroscopy (TDDS), all
  recovery events create exponentially distributed clusters on the
  spectral maps which do not move with increasing stress time.
\item[\mycnt] The location of these clusters is marked by a capture
  time, an emission time, and the step-height. In an agnostic manner,
  we also consider the forward and backward rates for the creation of
  interface states on the same footing.  The combination of such
  clusters forms a unique fingerprint for each nanoscale device.
\item[\mycnt] Given the strong bias- and temperature-dependence of the
  capture and emission times, the degradation in each device will have
  a unique temperature and bias dependence.
\end{itemize}
At the microscopic level, any BTI model describing charge trapping as well
as the creation of interface states should be consistent with the
above findings. Given the wide variety of published models, we have
compared two \emph{model classes} against these benchmarks, namely
\emph{reaction-} versus \emph{diffusion-}limited models.

\resetmycnt
As a representative for \emph{diffusion-}limited models, we have used
the \pRD{} reaction-diffusion model.
We have \emph{observed a complete lack of
  agreement}, as this non-dispersive reaction-diffusion model predicts
\mycnt{} that a very large number of equal interface states contribute equally to
recovery, while experimentally only a countable number of clusters can
be identified, \mycnt{} that the clusters observed in the spectral map
should be loglogistically distributed with an increasing mean value
given by the stress-time, and \mycnt{}
that the \emph{averaged} long-term degradation and recovery should be
roughly the same in all devices, independent of temperature and bias.
Based on these observations we conclude that the mainstream non-dispersive
reaction-diffusion models in their present form are unlikely to provide a correct
physical picture of NBTI.
These issues should be addressed in future variants of RD models and
benchmarked against the observations made here.

\resetmycnt
On the other hand, if we go to the other extreme and assume that NBTI
recovery is not \emph{diffusion-} but \emph{reaction}-limited,
the
characteristic experimental signatures are naturally reproduced.  Such 
models are \mycnt{} consistent with the exponential distributions in the spectral
map, \mycnt{} are based on widely dispersed capture and emission times which
result in fixed clusters on the spectral maps, and \mycnt{} naturally
result in a unique fingerprint for each device, as the parameters of
the reaction are drawn from a wide distribution.  As the time constants
are bias- and temperature-dependent, the unique behavior of each
device can be naturally explained and predicted, provided the
distribution of these time constants is understood.

Finally, we have argued that our results are not only interesting for
modeling enthusiasts, but have fundamental practical implications
regarding the way devices should be optimized and analyzed for reliability,
particularly for nanoscale devices, which will show increased
variability.

\section*{Acknowledgments} 

The research leading to these results has received funding from the
FWF project n$^\circ$23390-M24 and the European Community's FP7
project n$^\circ$261868 (MORDRED).  

\section*{Appendix}

\resetmycnt

In this Appendix three finer points are discussed,
namely \mycnt{} the subtle difference
between fully independent stress/relax cycles implied by \eq{e:RD-relax-pdf-single}
and a TDDS setting, \mycnt{} a possible impact of errors in
the discrete-step extraction algorithm, and \mycnt{} a contribution
of the quasi-permanent component.

\subsection*{Repeated Stress/Relax Cycles}

\begin{figure}[!t]
\begin{center}
\hbox{
    \includegraphics[width=\sfigwidth,angle=-90]{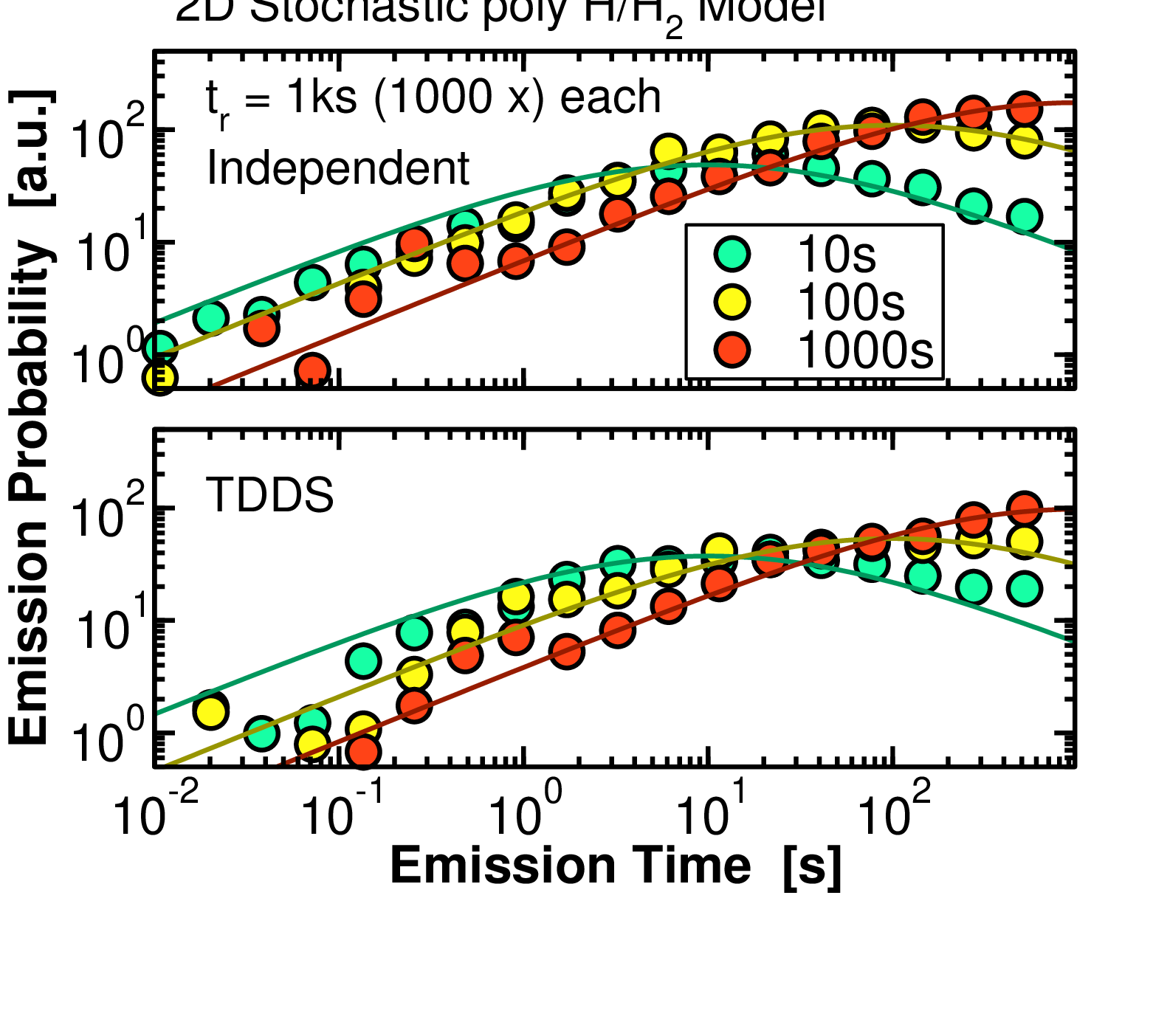}
    \hspace*{-4mm}
    \includegraphics[width=\sfigwidth,angle=-90]{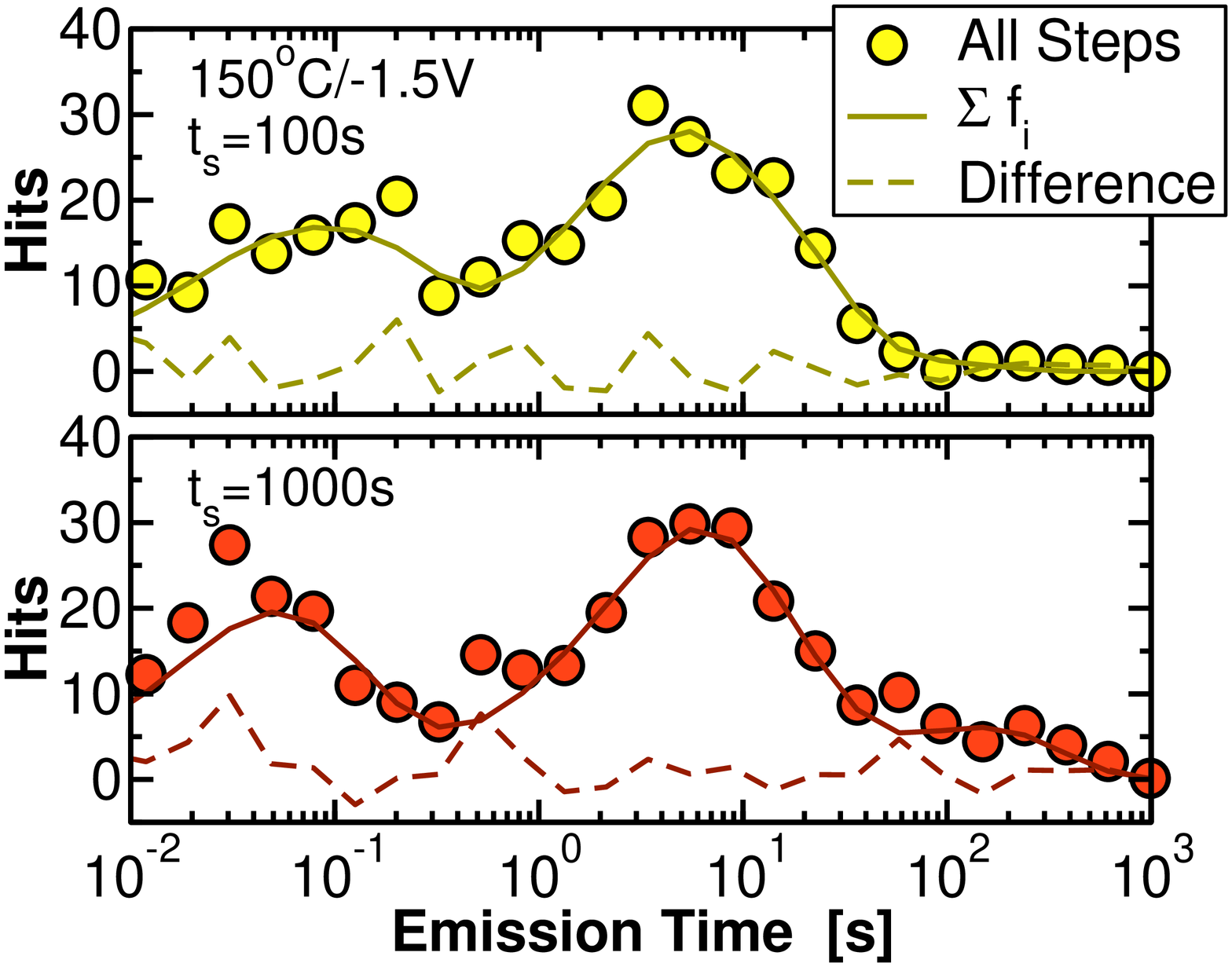}
}
    \vspace*{-6mm}

    \capspace\caption{\label{f:RD-Histo} \LFig: (Top) The emission probability predicted by the \pRD{} model (symbols are simulation) assuming independent stress/relax cycles follows a loglogistic
      distribution (lines). (Bottom) If the simulations are not conducted independently (that is, repeated on a completely recovered device) but in a TDDS setting (1000 repeated stress/relax cycles for better statistics),
      the number of emission events decreases but the statistics remain nearly unaffected.
\RFig: The sum of the exponential distributions fitted to the individual clusters (lines)
is subtracted from the total number of detected switches (symbols),
        revealing a certain noise in the data. However, no hidden RD component is identifiable in the noise.
}
\end{center}
\figspace
\end{figure}

Strictly speaking, equation \eq{e:RD-relax-pdf-single} is valid for a
single stress/relax cycle while the TDDS consists of a large number of
repeated cycles. As such, the TDDS setup corresponds to an
ultra-low-frequency AC stress and the devices will not be fully
recovered prior to the next stress phase.  This implies that \Hyd{}
would be able to move deeper into the gate stack during cycling and
that the \Hyd{} profile would not be precisely the same as that
predicted during DC stress \cite{ANG11}.  For short stress times and
long enough recovery times, e.g. \U{1}{s} versus \U{1}{ks}, the impact
of this would be small, since \eq{e:RD-relax-pdf-single} predicts
nearly full recovery in this case (97\%). However, for larger stress
times, recovery by the end of the cycle will only be partial and
\eq{e:RD-relax-pdf-single} may no longer be accurate in a TDDS
setting.  We have considered this case numerically in \Fig{f:RD-Histo} (left),
which shows that although this impacts the absolute number of recorded emission
events, the general features -- namely loglogistically distributed
clusters which move in time --  remains.

\subsection*{On Possible Extraction Errors}

As can be seen from \Fig{f:LongTermTDDS-Maps}, with increasing stress
time the number of visible clusters increases, as does the
noise-level, making an accurate extraction of the statistical
parameters more challenging than for shorter stress times.  In order
to guarantee that our extraction algorithm, which splits the recovery
trace into discrete steps, does not miss any essential features and
the noise in the spectral maps is really just unimportant noise rather
than an overshadowed RD contribution, we performed one additional
test:  we calculate the difference
between the extracted response of forward and backward reactions and
subtract it from all recorded steps, see \Fig{f:RD-Histo} (right).
As can be seen, even if due to noise not all steps are considered in
the fit, no hidden RD component is missed.

\subsection*{The Permanent Contribution to TDDS}

Finally, we comment on the permanent part that builds
up during the TDDS cycles, see \Fig{f:LongTermTDDS-Fit}.  This
contribution is not explicitly modeled here but only extracted from
the experimental data to be added to the modeled recoverable part.
From an agnostic perspective, one could simply refer to this permanent
build-up as due to those defects with emission or annealing times
larger than the maximum recovery time, \U{1}{ks} in our case.  This
permanent build-up is typically assigned to interface states (\Pb{}
centers) \cite{HUARD06}, but likely also contains a contribution from charge traps
with large time constants \cite{GRASSER11D}.

\black

\end{document}